\documentclass[usenatbib]{mn2e}
\usepackage{graphicx}
\def\simgt{\hbox{\rlap{\raise 0.425ex\hbox{$>$}}\lower 0.65ex\hbox{$\sim$}}}
\def\simlt{\hbox{\rlap{\raise 0.425ex\hbox{$<$}}\lower 0.65ex\hbox{$\sim$}}}
\def\arcsec{^{\prime\prime}}

\def\degree{^\circ}

\def\bj {b_{\rm J}}

\def\kms {km$\,$s$^{-1}$}

\def \oiii {[O{\small~III}]}

\def \sii {[S{\small~II}]}
\def \nii {[N{\small~II}]}

\def \ha {H$\alpha$}
\def \hb {H$\beta$}
\def \hi {H{\small~I}}

\def \ser {_{\rm ser}}
\def \re {R_{\rm e}}
\def \mue {\mu_{\rm e}}
\def \aj {AJ}
\def \mnras {MNRAS}
\def \apj {ApJ}
\def \apjs {ApJS}
\def \apjl {ApJL}
\def \aap {A\&A}
\def \pasp {PASP}
\def \pasj {PASJ}
\def \araa {ARA\&A}
\def \aapr {A\&AR}
\def \nar {newAR}
\def \fcp {FCPh}
\def \pasa {PASA}
\def \memsai {MmSAI}
\def \nat {Nature}
\title[The Sydney-AAO Multi-object IFS]{The Sydney-AAO Multi-object
  Integral field spectrograph (SAMI)}

\author[Croom et al.,]
{Scott M. Croom$^{1,2}$\thanks{scroom@physics.usyd.edu.au}, Jon S. Lawrence$^{3,4}$, Joss Bland-Hawthorn$^{1}$, Julia J. Bryant$^{1}$, 
\newauthor
Lisa Fogarty$^{1}$, Samuel Richards$^{1}$, Michael Goodwin$^{3}$,
Tony Farrell$^{3}$,
\newauthor
Stan Miziarski$^{3}$, Ron Heald$^{3}$, D. Heath Jones$^{5}$, Steve
Lee$^{3}$, Matthew Colless$^{3,2}$,
\newauthor
Sarah Brough$^{3}$, Andrew M. Hopkins$^{3,2}$, Amanda E. Bauer$^{3}$,
Michael N. Birchall$^{3}$,
\newauthor
Simon Ellis$^{3}$, Anthony Horton$^{3}$, Sergio Leon-Saval$^{1}$, Geraint Lewis$^{1}$,
\newauthor
\'A. R. L\'opez-S\'anchez$^{3,4}$, 
Seong-Sik Min$^{1}$, Christopher Trinh$^{1}$, Holly Trowland$^{1}$\\
${^1}$ Sydney Institute for Astronomy (SIfA), School of
Physics, University of Sydney, NSW 2006, Australia\\ 
${^2}$ ARC Centre of Excellence for All-sky Astrophysics (CAASTRO)\\
$^{3}$ Australian Astronomical Observatory, PO Box 296, Epping, NSW 1710,
Australia\\
$^{4}$ Department of Physics and Astronomy, Macquarie University, NSW 2109, Australia\\
$^{5}$ School of Physics, Monash University, Clayton, VIC 3800, Australia
}

\begin{document}

\maketitle

\newcommand{\fmmm}[1]{\mbox{$#1$}}
\newcommand{\scnd}{\mbox{\fmmm{''}\hskip-0.3em .}}
\newcommand{\scnp}{\mbox{\fmmm{''}}}

\begin{abstract}
  We demonstrate a novel technology that combines the power of the
  multi-object spectrograph with the spatial multiplex advantage
  of an integral field spectrograph (IFS). The Sydney-AAO Multi-object
  IFS (SAMI) is a prototype wide-field system at the
  Anglo-Australian Telescope (AAT) that allows 13 imaging fibre bundles
  (``hexabundles'') to be deployed over a 1--degree diameter field of view. Each
  hexabundle comprises 61 lightly--fused  multimode fibres with reduced
  cladding and yields a 75 percent filling factor.  Each fibre core
  diameter subtends 1.6 arcseconds on the sky and each hexabundle has
  a field of view of 15 arcseconds diameter. The fibres are fed to the
  flexible AAOmega double--beam spectrograph, which can be used at a
  range of spectral resolutions ($R=\lambda/\delta\lambda\approx$
  1700--13000) over the optical spectrum (3700--9500\,\AA). We present
  the first spectroscopic results obtained with SAMI for a sample of galaxies at $z \approx
  0.05$. We discuss the prospects of implementing hexabundles at a
  much higher multiplex over wider fields of view in order to carry
  out spatially--resolved spectroscopic surveys of $10^4-10^5$ galaxies.

\end{abstract}


\section{Introduction}
\label{sec:intro}

Galaxies are intrinsically complex with multiple components and varied
formation histories. This complexity is the primary reason that
unravelling the physics of galaxy formation and evolution is so
challenging. Galaxies are made up of baryons confined to dark matter
haloes, and often have multiple distinct kinematic components (e.g.\
bulge and/or disc). There are complex interactions between the stars,
gas, dust, dark matter and super-massive black holes. These can lead to
both positive and negative feedback on the formation rate of stars.

Experimental efforts to address galaxy formation have generally taken
two directions. First, galaxy imaging and spectroscopic surveys have
progressively moved to higher redshift, in an attempt to directly
observe galaxy evolution and formation. This approach has had much
success, placing strong constraints on the evolution of the global star
formation rate \citep[e.g.][]{2006ApJ...651..142H}, unveiling the strong
evolution of black hole accretion over most of cosmic time
\citep[e.g.][]{2004MNRAS.349.1397C,2009MNRAS.399.1755C,2006AJ....131.2766R},
tracing the evolution of galaxy size and morphology
\citep[e.g.][]{1997ApJ...490..577D} and much more.

The second approach has been to expand our view in wavelength rather
than cosmic time. The physical processes occurring in galaxies cause
emission over the entire range of the electromagnetic spectrum. In order
to have a full picture of galaxy formation, a multi-wavelength approach
is vital. This has been made more achievable with recent generations of
satellites covering the X-ray, ultra-violet, mid- and far-infrared.
While the spectral energy distributions of stars tend to peak in the
optical or near-infrared, obscuration and reprocessing by dust generates
strong mid- and far-infrared emission. Young stars (when not obscured)
are most directly traced in the ultraviolet, while black hole accretion
can generate radiation from the radio to X-ray and gamma-ray bands.
Surveys such as the Galaxy And Mass Assembly (GAMA) survey
\citep{2011MNRAS.413..971D} and the Cosmic Evolution Survey
\citep[COSMOS;][]{2007ApJS..172....1S} show the value of this
multi-wavelength approach.

The third route, and the one that we address in this paper, is to focus
on spatially resolving galaxies; in particular, obtaining spatially
resolved spectroscopy. Optical spectroscopy allows us to measure a wide
range of parameters including current star formation rates (e.g.\ via
\ha), gas phase metallicities, stellar ages, stellar metallicities,
black hole accretion rates, ionization structure and extinction due to dust
(e.g.\ via the Balmer decrement). The major spectroscopic surveys to
date have used a single fibre
\citep{2001MNRAS.328.1039C,2003AJ....126.2081A} or single slit
\citep{2005A&A...439..845L,2007ApJ...660L...1D} on each object, and so
obtain just one measurement of these parameters for each galaxy.
Moreover, these measurements may not be representative of the galaxy as
a whole, but biased, depending on where the aperture
is placed. This fundamental problem is addressed by integral field
spectrographs (IFS). In the last decade, projects such as SAURON
\citep{2001MNRAS.326...23B} have demonstrated the power of integral
field spectroscopy to capture a range of key observables that are simply
not available to single-aperture surveys. As well as studying the
properties listed above in a spatially-resolved context, obtaining gas
and stellar kinematics over an entire galaxy enables us to separate
dynamical components, measure dynamical mass, examine the impact of
winds and outflows, and discover merging systems via dynamical
disturbances.

Integral field spectroscopy has almost exclusively been limited to
single-object instruments, meaning that it is time-consuming to build
large samples. The largest current data set, using the SAURON system on
the William Herschel Telescope, contain $\sim260$ objects
\citep[ATLAS-3D;][]{2011MNRAS.413..813C}. The CALIFA project
\citep{2011califa} aims to target 600 objects with the PMAS integral
field unit (IFU)
on the Calar Alto Telescope using $\sim$200 nights of telescope time.
The only multi-object integral field spectrograph currently available is
that on the VLT FLAMES instrument \citep{2002Msngr.110....1P}, which
contains 15 IFUs, each with 20 spatial elements
of size 0.52 arcsec and a total field of view (FoV) of $2\times3$\,arcsec.
This has enabled measurements of the Tully-Fisher relation at $\sim$0.6
\citep{2008A&A...477..789Y} as well as a number of other projects.
However the small FoV of each IFU, combined with the high
spectral resolution ($\ge$9000) and associated narrow wavelength range
limits its applicability for large--scale surveys.

Astrophotonic technology \citep{2009OExpr..17.1880B} is now opening the
way to new instrumentation that can address the need for highly
multiplexed integral field spectroscopy. Hexabundles
\citep{2011OExpr..19.2649B,2011MNRAS.415.2173B} are optical fibre
bundles where the cladding has been stripped from each fibre to a
minimum over a short length ($\sim$30\,mm) and the fibres then gently
fused together at the input end to provide an IFU ($\sim$1\,mm
aperture) with high filling factor. These can then be used in
conventional multi-fibre spectrographs.

In this paper, we report on the Sydney-AAO Multi-object Integral field
spectrograph (SAMI),
the first fully operational demonstrator instrument to use
hexabundles. 
SAMI has 13 IFUs that can be positioned anywhere over a 1 degree
diameter field
of view. In Section~\ref{sec:rationale}, we discuss in detail the
scientific rationale for such an instrument, along with some practical
considerations regarding sensitivity. In Section~\ref{sec:instrument},
we describe the SAMI instrument in detail. In Section~\ref{sec:comm}, we
outline the observations carried out during the commissioning of the
instrument, the results from which are discussed in
Section~\ref{sec:results}. Our conclusions, and goals for the future,
are laid out in Section~\ref{sec:conc}.

\section{Scientific Rationale}
\label{sec:rationale}

In this section we outline the key scientific drivers for an instrument
such as SAMI. The fundamental question at the heart of this work is: how
did the galaxy population we see around us today come about? This
requires us to understand the physical processes that occur as galaxies
form and evolve. The galaxy population we see today has some very
distinctive features that need to be explained. 

One of the most fundamental is the separation of galaxies into a bimodal
distribution according to colour
\citep[e.g.][]{2001AJ....122.1861S,2004ApJ...600..681B}. A galaxy's
colour is primarily related to the presence of ongoing star formation.
The second key feature differentiating galaxies is morphology. There is
a strong correlation between colour and morphology, with galaxies lying
along the `red sequence' being mostly passive systems with
elliptical/spheroidal morphology, while galaxies inhabiting the `blue
cloud' are mostly dominated by discs, although this is not always
strictly the case
\citep[e.g.][]{2010MNRAS.405..783M,2009MNRAS.396..818S}. While they are
related, there is not a strict one-to-one relationship between
morphology and colour. A clearer understanding of galaxies can be
obtained if they are considered as being made up of distinct
morphological components (discs, bulges and pseudo-bulges) that result
from different formation processes and evolutionary histories
\citep{2007ApJ...657L..85D}. The intrinsic properties of these
structural components are more uniform than those of the galaxies they
compose. Their formation pathways are also quite different, with true
bulges built up by violent mergers, discs from gas accretion, and
pseudo-bulges from secular evolution \citep{2004ARA&A..42..603K}.
Disentangling these various modes is complex, but can be materially
aided by the fact that the structures formed have different kinematic
properties as well as different star formation histories.

In order to ascertain the physics involved, we need to determine the
answers to a number of questions that broadly fall into four categories:
(i)~how does galaxy mass and angular momentum build up? (ii)~when, where
and why does star formation occur? (iii)~when, where and why does black
hole accretion occur? (iv)~how are galaxies fuelled and what is the role
of feedback? We will discuss each of these issues in turn, although
there is of course significant overlap between them.

\subsection{The build up of mass and angular momentum}

The standard picture of galaxy formation has gas cooling to form a
rotationally supported disc within a cold dark matter halo
\citep{1978MNRAS.183..341W}.  While this picture is broadly accepted,
feedback and interactions provide major complications which are not
yet fully understood.    

The scaling relation between circular velocity and stellar mass
\citep[The Tully-Fisher Relation, TFR;][]{1977A&A....54..661T}
for disc galaxies provides a tight constraint for galaxy formation
models.  The circular velocity depends on the ratio of disc mass to
halo mass, the dark matter halo profile and the dimensionless spin
parameter, $\lambda$ \citep{1969ApJ...155..393P}.  The largest  TFR
samples \citep[e.g.][]{2007ApJS..172..599S} currently contain
$\simeq5000$ galaxies with either long-slit spectroscopy or spatially
unresolved \hi\ velocities.  If IFU observations can be made out to
large enough radii (typically $\sim2.2$ disc scale lengths), then they
provide substantial advantages in allowing a clearer picture of
distorted kinematics and inclination.  Circular velocities can be
compared to the results of galaxy lensing to constrain the dark matter
halo profile and examine evidence of contraction of the halo in response
to the baryons \citep{2010MNRAS.407....2D,2011arXiv1110.4107R}.    

The stellar and emission line kinematic data that SAMI can
provide will allow dynamical mass estimates within maximum radius
probed, using techniques such as anisotropic Jeans modelling
\citep{2008MNRAS.390...71C}.  In general it is not possible to
determine total mass because of the uncertainty of the dark matter halo
parameters. Even for the Milky Way, where many halo stars can be used to
probe the outer halo, the total galaxy mass is uncertain to a factor of
two \citep{2007MNRAS.379..755S}.  Detailed dynamical techniques are in
stark contrast to estimates of dynamical mass which simply take a single
velocity dispersion of the galaxy (e.g. from a single fibre
observation), \citep[e.g.][]{2010ApJ...722....1T}.

It is relatively easy to extract a rotation curve $v(r)$ from the
observed data if the kinematics are fairly well ordered
\citep{1990ApJ...364...23S}.  This can then be used to provide dynamical
information about the galaxy, particularly if baryonic information is
brought to bear. However if the kinematic axes are misaligned with the photometric axes,
this is often a signature of streaming motions due to a bar or an oval
distortion. A dynamical mass can still be derived, although the
increased number of free parameters makes this more uncertain
\citep{1990ApJ...364...23S,1994ApJ...437..162Q}. Any deviations from
rotational symmetry are important in their own right. It is often very
difficult to see the presence of bars, particularly in highly-inclined
disc systems. But bars are often betrayed by the inner twists of the
isovelocity contours. Warps are more easily detected on large scales and
almost always in HI kinematics \citep{1990ApJ...352...15B}; however, the
same effects can now be seen in deep observations of the diffuse ionized
gas in the outer disc \citep{2010MNRAS.405.2549C}.  The physical cause
of kinematic distortions can be examined by large IFU surveys which
probe a variety of galaxy parameters.  For example, are distortions
more likely in high density regions due to dynamical interaction.

Integral field spectroscopy also enables studies of stellar kinematics
that describe the observed projected stellar angular momentum per unit
mass of galaxies, not possible with other techniques
\citep{2007MNRAS.379..401E,2011MNRAS.414..888E,2011MNRAS.414L..80B}. 
This measurement enables the separation of early-type galaxies into
fast and slow rotators which are thought to have very different
formation paths.  A great success of the SAURON and ATLAS-3D projects
\citep{2001MNRAS.326...23B,2011MNRAS.413..813C} has been the discovery
that most early type galaxies have significant rotation, with only
$\simeq14$ percent (predominantly at high mass) being slow rotators.
\citet{2011MNRAS.416.1680C} have used ATLAS-3D to demonstrate a
kinematic morphology-density relation, which shows a smooth transition
of spirals to early type fast rotators with increased density, and
massive slow rotators only inhabiting the highest density regions.
SAMI would allow such studies to be expanded to probe greater dynamic
range in environment and examine a such relations as a function of
stellar mass.

The rate of dark matter halo merging can be accurately estimated from
simulations \citep{2008MNRAS.386..577F}.  Kinematic information from
integral field spectroscopy can be used to differentiate between
quiescent galaxies and those undergoing a merger
\citep[e.g.][]{2008ApJ...682..231S}, using procedures such as
kinemetry \citep{2006MNRAS.366..787K}.  Until now, this type of analysis has
largely been limited to high redshift (where the merger rate is
expected to be higher), but with samples of $10^3$ or more galaxies, statistically
meaningful  estimates of merger rates can be made at low redshift.
\citet{2008MNRAS.386..577F} predict that the halo merger rate at
$z\sim0$ should be $\sim0.05$\,halo$^{-1}$\,Gyr\,$^{-1}$ for major
mergers (with mass ratios $<3/1$).   Integral field observations
provide a complementary approach to studies which focus on the number
of close pairs to estimate merger rates
\citep[e.g.][]{2007ApJ...666..212D}, as they probe very different
phases of the merger process.  It is also be possible to look
for weaker dynamical disturbances in discs due to repeated minor
mergers/interactions \citep{1997ApJ...477..118Z}.  In this case, it is
very advantageous to extend the study to large radius as the effect of
tidal perturbations scales as $\sim r^3$.  In this case the challenge
is to achieve sufficient sensitivity at large radius
(e.g. $\simgt1.5-2$ scale lengths).

Spin angular momentum from galaxy kinematics can directly probe the
formation of the large scale structure of the universe and galaxy
formation. The spin of galaxy discs provides an approximation of the
spin of the galaxy's dark matter halo \citep{2005ApJ...628...21S}, which
is coupled to the large scale structure. Early in a dark matter halo's
life, it experiences torques from the surrounding density landscape. The
spins of dark matter haloes today retain some memory of that landscape,
so spin is intrinsically linked with the large scale structure. This
link may be examined in N-body simulations and observations by measuring
the distribution of inferred dark matter halo spin magnitude
\citep{2008MNRAS.391..197B} and by the relative orientation of galaxy
spin directions with each other and with the large scale structure.

N-body simulations do not predict a strong alignment between the spins
of neighbouring haloes, although an alignment between galaxies has been
detected in the Tully catalogue of 12,122 nearby spirals
\citep{2000ApJ...543L.107P}. Simulations and theory predict an alignment
between halo spin and the tidal field \citep{2000ApJ...532L...5L}, and
an alignment with features in the tidal field like filaments
\citep{2009ApJ...706..747Z}, sheets \citep{2004ApJ...614L...1L} and
voids \citep{2007MNRAS.375..184B}. Any kind of alignment is
predicted to be very weak, however, so could only be seen in large scale galaxy
surveys. There have been detections of spin alignments in the large
scale structure reconstructed from imaging surveys
\citep{2007ApJ...671.1248L,2008MNRAS.389.1127P}, using inferred galaxy
spins from disc shapes. A tentative detection of spin alignment with filaments was found
using the inferred spin of only 201 galaxies around voids
\citep{2006ApJ...640L.111T} and 70 galaxies in filaments
\citep{2010MNRAS.408..897J}, picked from the large scale structure of
SDSS. Discrepancies between the results found from dark matter
simulations and observations indicate a difference in the way that
galaxies and dark matter haloes obtain and keep their spin. A large
survey of direct spin measurements could reveal whether galaxies exhibit the
same spin behaviour as dark matter haloes, and show how galaxy spin is
linked to the large scale structure.

\subsection{When, where and why does star formation occur?}

Much recent observational and theoretical work has focussed on how blue
galaxies can have their star formation quenched, moving them onto the
red sequence. Red sequence galaxies are preferentially found in denser
environments \citep[e.g.][]{2005ApJ...629..143B}, and star formation is
also clearly suppressed at high density
\citep[e.g.][]{2002MNRAS.334..673L}. This immediately suggests
environmental factors play an important role. When a galaxy falls into a
cluster, the {\it ram pressure} from the dense intergalactic medium
\citep{1972ApJ...176....1G} may expel the gas from the disc, removing
the fuel required for further star formation. There are several observed
examples of this in rich clusters
\citep[e.g.][]{2008ApJ...688..208R,2007ApJ...671..190S}. In moderately
dense regions, such as galaxy groups, it may be that ram pressure will
leave the disc intact, but can still remove gas from the halo of the
galaxy. The halo provides a reservoir of gas which can replenish the
disc.

Without the halo gas, the star formation will decline and then cease as
the disc gas is consumed, leading to a transition to the red sequence,
in a process known as {\it strangulation} \citep{1980ApJ...237..692L}.
Simulations suggest that this process can be efficient at removing halo
gas, even in small and/or compact groups
\citep{2009MNRAS.399.2221B,2008MNRAS.383..593M}, but there is little
direct experimental evidence that this is the case. Indirect evidence
does point to some pre-processing of galaxies in groups before they fall
into clusters \citep{2010MNRAS.402L..59B}, but the physical process
driving this is not clear. Direct galaxy-galaxy interactions are also
expected to play a critical role, with major galaxy mergers triggering
star formation \citep[e.g.][]{2008AJ....135.1877E} and transforming the
morphology of galaxies, although the fraction of galaxies undergoing
major mergers (i.e.\ those with mass ratios of 3:1 or less) in the local
Universe is small \citep[e.g.][]{2008ApJ...685..235P}.  On the other
hand, dwarf star-forming galaxies in the local Universe are often
found interacting with low-luminosity objects or diffuse H{~\small I}
clouds,.  This appears to be the triggering mechanism of their intense
star-formation activity
\citep{2008A&A...491..131L,2009A&A...508..615L}, although only
detailed multi-wavelength observations are able to reveal these processes
\citep{2010A&A...521A..63L,2011arXiv1109.0806L} 

However, environment is only one factor. Feedback from star formation
and accretion onto super-massive black holes provides an internal
mechanism for transformation.  This feedback provides a solution
to the mismatch of the theoretical dark matter halo mass function and
the observed stellar mass function \citep[e.g.][]{2008MNRAS.388..945B} by
heating and/or expelling gas in both low mass (via star formation) and
high mass (via black hole accretion) haloes
\citep{2006MNRAS.370.1651C,2008MNRAS.388..945B}. Extreme outbursts of star
formation or black hole accretion may be triggered by mergers or
interactions \citep[e.g.][]{2008ApJS..175..356H}, making a link between
internal and environmental effects. Once the burst is over, another
mechanism is needed to suppress continued star formation. The best
suggestion for this is mechanical feedback from jets emitted by
super-massive black holes \citep[e.g.][]{2006MNRAS.365...11C}, but this only
appears to be efficient in massive galaxies.

As well as these active processes, the environment has an indirect
influence via formation age. Galaxies in high density regions form
earlier and so have had more time to evolve
\citep[e.g.][]{1984ApJ...284L...9K,1986ApJ...304...15B,2005ApJ...621..673T}. In
the absence of other effects, we would then expect to see galaxies in high
density regions having older stellar populations.

Disentangling these varied influences on galaxy formation is far from
trivial. However, studies of the spatial distribution of instantaneous
star formation rates, integrated star formation (via stellar population
ages) and metallicity (both gas and stellar) provide considerable
insight. Importantly, ram-pressure removal of gas implies that the
truncation of star formation is an {\it outside-in} process
\citep[e.g.][]{2009MNRAS.399.2221B,2009A&A...499...87K}. Gas is
preferentially removed in the outer
parts of galaxies, which are less gravitationally bound. This may be a
short-lived feature of the galaxies in dense environments if the gas is
eventually completely removed. Alternatively, stripping can occur over
the lifetime of a galaxy if the gas is puffed up by the internal star
formation; in this case, even a rarified external medium can remove gas
from the galaxy \citep{2011ApJ...732...17N}. Globally, the
expectation would be that galaxies form inside-out, and this implies age
and metallicity gradients which are observed
\citep[e.g.][]{1983MNRAS.204...53S,1992MNRAS.259..121V,1994A&A...281L..97S,1997ApJ...477..765C}.

One approach that has been explored in some detail as an alternative to
spatially-resolved spectroscopy is the so-called `pixel-z' technique
\citep{2003AJ....126.2330C,2009ApJ...701..994W,2008ApJ...677..970W}.
This approach, analogous to photometric redshifts (`photo-z'), uses a
library of template spectral energy distributions (SEDs) to fit the
observed optical and near-infrared colours of individual pixels within
resolved galaxy images. This technique has been used with some success
to explore the environmental dependence of star formation in galaxies.
\citet{2008ApJ...677..970W} find that, globally, the suppression of star
formation in high density environments
\citep[e.g.,][]{2002MNRAS.334..673L,2003ApJ...584..210G} seems to occur
primarily in the most strongly star-forming population, and to be
evidenced by a suppression in the {\em inner} regions of galaxies.
\citet{2009ApJ...701..994W} demonstrate that this effect seems to hold
independently for both early- and late-type galaxy populations, and that
the suppression in star formation cannot be explained solely by the
well-known density-morphology relation
\citep[e.g.,][]{1980ApJ...236..351D}. There are significant limitations,
however, to the `pixel-z' approach. These are related to implicit
assumptions made by the technique, such as each pixel being represented
by an isolated single stellar population with a simple exponential
star-formation history \citep{2003AJ....126.2330C,2008ApJ...677..970W}.
As a result, the method cannot effectively measure the instantaneous
star formation rate, which can be traced spectroscopically (e.g.\ by
\ha\ emission).

IFU spectroscopy allows us access to both current star formation (via
emission lines) and integrated star formation history (via stellar age
and metallicity). Examining the radial dependence of the mean stellar
age and metallicity gradients tells us when and where the stars formed
in these galaxies, along with a fossil record of the galaxy merger
history. The mean stellar age is effectively a luminosity-weighted
integral of the star formation history whilst the stellar metallicity
gradient provides an indication of its merging history
\citep[e.g.][]{2004MNRAS.347..740K}. If environmental effects are
responsible for the cessation of 
star formation, then we would expect red sequence galaxies to have
younger central ages with past major mergers sign-posted by shallow
negative metallicity gradients \citep[i.e. lower metallicities in the
outskirts;][]{2004MNRAS.347..740K,2007MNRAS.378.1507B,2009ApJ...691L.138S}.

\subsection{When, where and why does black hole accretion  occur?}

A full picture of the physical processes involved in the fuelling of accretion onto
super-massive black holes, resulting in the phenomenon of an active
galactic nucleus (AGN), is still elusive.  
It is now known that most galaxies contain super-massive black holes,
with typical masses a million to a billion times that of the Sun
\citep[e.g.][]{2000ApJ...539L..13G,2000ApJ...539L...9F}. The mass of
the black hole 
correlates well with the mass (or velocity dispersion) of the bulge or
spheroidal component \citep{2002ApJ...574..740T} in a galaxy. This implies an
intimate connection between the build up of stellar mass in galaxies and
their super-massive black holes. 

The nature of the connection between star formation and AGN has long
been debated \citep[e.g.][]{1988ApJ...325...74S}, although a resolution remains
elusive. The most luminous AGN (i.e.\ bright quasars) require
$\simgt$10$^9$ solar masses of gas deposited in their central regions on
time-scales of $\sim$10$^7$--10$^8$ years \citep[e.g.][]{2005MNRAS.356..415C},
requiring major galaxy-wide perturbations \citep{2009ApJ...694..599H},
and no doubt triggering substantial star formation. In contrast,
low-luminosity AGN require relatively small amounts of gas, which can be
supplied by internal stochastic processes, such as the accretion of cold
molecular clouds \citep{2009ApJ...694..599H}.

The AGN population evolves strongly with cosmic time.  This is most
clearly seen in evolution of luminous quasars from $z=0$ to $z\simeq6$
\citep[]{2006AJ....131.2766R,2009MNRAS.399.1755C}, which have a strong
peak in space density at $z\simeq2-3$.  Lower luminosity AGN show less
severe evolution, and peak in space density at lower redshift
\citep[e.g.][]{2005A&A...441..417H,2009MNRAS.399.1755C}.  This
downsizing appears qualitatively similar to that seen in the
formation of galaxies \citep{1996AJ....112..839C}.  In the local
Universe, accretion onto black holes is dominated by systems with low 
black hole masses, $<10^8$\,M$_\odot$, but which are typically
accreting within an order of magnitude of the Eddington limit
\citep{2004ApJ...613..109H}.  In contrast, higher mass black holes in
the local Universe are accreting with a characteristic timescale
substantially longer than a Hubble time.

Luminous AGN (as measured by their \oiii\,$\lambda$5007 luminosity)
are found to have younger stellar populations
\citep{2003MNRAS.346.1055K}, as measured by their $D_{\rm n}$(4000) and
H$\delta$ line indices from SDSS spectra.  It is not clear whether
this younger stellar population is centrally concentrated, or is
distributed more widely across the host galaxy, as the SDSS fibres
subtend a physical scale of $\sim5$ kpc at $z=0.1$.
IFS data has the ability to explicitly examine the distribution of
star formation and stellar population ages across galaxies and
investigate whether these are related to accretion rate.  

Lower luminosity AGN tend to have spectra typical of Low-Ionization Nuclear
Emission-line Regions \citep[LINERS;][]{1980A&A....87..152H}.  While
there is evidence that these form a continuous sequence with Seyferts
\citep{2006MNRAS.372..961K}, there substantial evidence that LINER like
emission is extended and may be powered by ionization from AGB stars
\citep{2011arXiv1109.1280Y}.  A large galaxy survey with spatially
resolved spectroscopy could resolve this issue.  If LINER emission is
not due to a central AGN in most cases this would require substantial
re-interpretation of recent work on low redshift AGN.
At an even more fundamental level, the fraction of galaxies which host
AGN is only well defined locally \citep[i.e. 10s of
Mpc;][]{2008ARA&A..46..475H}.  Approximately $40$ per cent of these very local
galaxies host AGN.  At larger distances contamination from off nuclear
emission increasingly reduces the sensitivity to weak nuclear emission
lines.  A large IFS survey would enable apertures of fixed metric size
to be defined, enabling robust AGN rates as a function of redshift to be determined.

While there is a good theoretical basis for galaxy mergers triggering
luminous AGN \citep[e.g.][]{2009ApJ...694..599H}, the evidence for this is
mixed.  An alternative pathway is via violent disc instability in self
gravitating discs
\citep[e.g.][]{2006MNRAS.370..645B,2011arXiv1107.1483B}.  Disc
instability is at least likely to play a role at high redshift where
cold streams can dominate the mass accretion onto discs
\citep{2009Natur.457..451D}.  Indeed, a large fraction of high--redshift star forming
galaxies appear to be discs
\citep[e.g.][]{2008ApJ...687...59G,2011MNRAS.tmp.1511W}, rather than mergers.  In low redshift samples AGN activity is not
enhanced by the presence of a nearby companion
\citep{2008MNRAS.385.1915L}, while star formation is
\citep{2008MNRAS.385.1903L}.  This is a somewhat surprising result, and
may point to a difference in timescale between the onset of star
formation and the AGN, with the AGN occurring later, after the merger
has taken place.  Indeed, IFS observations of local galaxies with observed
outflows demonstrates that the AGN timescale is significantly longer
than the starburst timescale \citep{2010ApJ...711..818S}.  Large scale
IFS observations can directly address the issue of AGN fuelling by
examining the kinematic properties of AGN hosts, and searching for
evidence of disc instability and/or merging.  In this regard it will be
particularly important to span a range of accretion luminosities in
order to examine whether there is a change from secular evolution at
low accretion rate to mergers at high accretion rate.

\subsection{Feeding and feedback}

The accretion of gas onto galaxies remains a largely unsolved problem.
Whether gas enters the galaxy potential in a hot phase and then cools
down \citep{2009MNRAS.397.1804B}, as a warm rain
\citep{2007ApJ...670L.109B}, or as an HI complex like the high velocity
clouds in the Galactic halo \citep{2008A&ARv..15..189S}, or all of the
above, is unclear. After a review of the evidence,
\citet{1992ARA&A..30...51B} concluded that the outer warps of HI discs
were some of the best evidence of ongoing disc accretion. Once the gas
settles into the galaxy, the outstanding issues are how the gas feeds
into the nuclear regions and, in particular, onto a central black hole.

In a survey of $10^3-10^4$ galaxies, SAMI offers the prospect of studying
nuclear activity (AGN, starburst, LINER) and star formation within the
context of the extended galaxy. The kinematic signatures of outer warps
and inner bar streaming are relatively easy to pick out in HI
\citep{1990ApJ...364...23S} or in ionized gas
\citep{2010MNRAS.405.2549C}. Thus, we can now directly associate this
activity with large-scale disc disturbances, assuming these exist.
Traditionally, in large galaxy surveys, the association of activity
and dynamical disturbances is made from the proximity of 
galaxies in position and redshift space
\citep[e.g.][]{2004MNRAS.355..874N,2008MNRAS.385.1903L}.  With full kinematic information
a much more direct determination of the triggering of activity will be
possible. 

An alternative approach is to study the impact of the inner disc on the
extended properties of galaxies
\citep[e.g.][]{1998ApJ...506..222M,2003MNRAS.345..657K}. Jets carry
energy, and winds carry gas and metals, far from the nucleus. In a
recent integral field study of ten AGN and starburst galaxies,
\citet{2010ApJ...711..818S} find that starburst winds are largely shock
ionized, while AGN winds show the hallmark of photo-ionization by the
accretion disc, clearly indicating that the starburst phenomenon is very
short. For one of the objects observed on the SAMI
commissioning run, we see the ionization characteristics typical of
nuclear activity for gas off the plane of an inclined disc (see
Section \ref{sec:gals} and Fogarty et
al., in prep.). A large-scale wind is confirmed by the broad
emission-line profiles along the minor axis. This is a remarkable
testament to the power of spatially-resolved kinematic and ionization
information. In the full survey, SAMI is likely to uncover hundreds of
new outflow sources connected either to nuclear activity or inner disc
star formation. There is an even rarer class of galaxies with disc-wide
winds that SAMI will also inevitably add to \citep{2007MNRAS.376..523S}.

A number of edge-on spiral disc galaxies have vertically extended
ionized gas in their haloes \citep{1996ApJ...462..712R,1992FCPh...15..143D}. The Reynolds Layer
in our Galaxy, recently mapped by the WHAM \ha\ survey telescope \citep{2006MmSAI..77.1163M}, is a good example of this phenomenon \citep[see also][]{2008PASA...25..184G}. The ionization characteristics of this gas does
not appear to be consistent with any known mechanism (i.e.\ UV
photo-ionization by hot young stars, radiation from old supernova
bubbles, shocks from supernovae, cosmic ray heating, or radiation
pressure on dust grains in the disc). The gas may arise from some kind
of disc-wide interaction between the disc and the hot halo, presumably
driven by processes related to star formation in the disc \citep{2005ARA&A..43..337C}.
But there is also the prospect that some of this gas is related to warm
gas accretion onto the disc \citep{2009IAUS..254..241B} or involved in a
large-scale circulation or recycling of gas through the halo \citep{2011MNRAS.415.1534M}.

As with the study of galaxy winds, a large SAMI survey has the
potential to greatly increase the sample of known galaxies with
vertically-extended warm discs. With a larger sample, it will be
possible to correlate the presence of these discs with the disc star
formation rate, nuclear activity and galaxy mass.

\subsection{Limitations of  current spectroscopic surveys}

Historically, telescopes were used to observe one source at a time. But
with technical advances in optical fibres, it was realized in the early
1980s that many sources could be observed simultaneously across the
telescope focal plane by precisely positioning fibres in the field
\citep{1981PASP...93..154B,1983SPIE..374..160G}. This led to an
explosion in wide-field spectroscopic surveys, notably including the 2-degree Field Galaxy Redshift
Survey 
\citep[2dFGRS;][]{2001MNRAS.328.1039C}, 2-degree Field QSO Survey
\citep[2QZ;][]{2004MNRAS.349.1397C}, 6-degree Field Galaxy Survey
\citep[6dFGS;][]{2009MNRAS.399..683J} and  Sloan Digital Sky Survey
\citep[SDSS;][]{2000AJ....120.1579Y} amongst several others. Between them, such surveys have 
obtained spectra for approximately 1.5 million extragalactic targets. New
instruments recently commissioned
\citep[e.g. LAMOST;][]{1998SPIE.3352...76S} or in construction
\citep[e.g. VIRUS;][]{2004SPIE.5492..251H} are able to observe
thousands of sources at a time.

Considerable advances have been made possible by the 2dFGRS and SDSS,
which use a single optical fibre 
per galaxy. However, with fibre diameters of 2 and 3 arcseconds
respectively, these projects sample less than half the light from a
galaxy at the median distance of the surveys. These single apertures
limit the surveys in two ways.

First, it is impossible for single-fibre surveys to measure spatially
varying spectral properties, which prohibits the study of crucial
observables such as kinematic merger rates, galaxy rotation and
dynamical mass, star formation gradients, metallicity gradients, age
gradients and detection of galaxy winds and/or outflows.

The second limitation is that with single-fibre spectroscopy the
measured signal depends on many things: (i)~intrinsic properties, like
source luminosity, size and distance; (ii)~atmospheric conditions,
particularly seeing; (iii)~instrumental properties, like fibre aperture
size and positioning accuracy, and optical focus over the field;
(iv)~telescope properties, such as pointing and guiding precision; and
perhaps other effects. Many published papers make the mistake of
assuming the surveys provide {\it spectrobolometry} (i.e.\ the spectrum
of the total light output by the source) rather than the spectrum from
an (often ill-defined) spatial sample of the source. 
The inherent dangers of aperture effects have long been known in
astronomy, but have often been under-appreciated or ignored.  \citet[see their Fig. 8]{2005MNRAS.363.1257E} clearly demonstrate
that the single-aperture fibre spectra from a typical galaxy survey may
be only weakly correlated with the photometric classification.

Aperture biases can manifest themselves in two ways. The fibres subtend
an increasing linear size with increasing distance of the galaxy,
potentially causing spurious evolutionary effects (and spurious
luminosity-dependent effects in a flux-limited sample). Second,
important galaxy properties, such as star formation rate and metallicity
can have substantial gradients, meaning that observations of just the
central regions are not representative of global values \citep[e.g.][]{2008ApJ...681.1183K}.

\subsection{Why multiplexed IFUs?}

Integral field spectroscopy allows us to gather data on key observables
that are simply inaccessible to single apertures. IFUs are usually a single monolithic array of lenslets, with the light
fed to a spectrograph via optical fibres (or alternatively an optical
image slicer). As a consequence, they can only target one object at a
time and so IFU surveys have typically only targeted a few tens of
galaxies in specific classes \citep[e.g.][]{2009MNRAS.396.1349P,2007MNRAS.379..401E}.

The process of galaxy formation and evolution is inherently complex,
with the observed galaxy properties depending on a large number of
parameters, such as host halo mass, stellar mass, merger history etc. As
well as the multi-dimensional nature of galaxy properties, there is
inherent stochasticity in the process. This is at least in part due to
our inability to accurately trace the formation history of individual
objects, but also derives from inherently non-linear physics such as
that involved in the collapse of molecular clouds to form stars.

The multi-dimensional nature of the galaxy population, combined with
this stochasticity (which at some level could be considered as extra
hidden parameters), means that large surveys are required to extract the
key relations between physical properties. The success of surveys such
as 2dFGRS and SDSS has in large part been due to their ability to `slice
and dice' the galaxy distribution and still have statistically
meaningful samples in each bin of the parameter space.

The need to extend integral field observations to large samples is well
understood in the above context, and has driven recent projects such as
ATLAS-3D \citep{2011MNRAS.413..813C} and CALIFA \citep{2011califa}.
These impressive projects are limited by their use of monolithic
integral field units, which means that targeting more than a few
hundred objects is prohibitively expensive in terms of telescope time.
This limitation naturally drives us to multi-object integral field
spectroscopy (i.e.\ multiplexed IFUs), the subject of this
paper.

\subsection{Size and surface brightness}\label{sec:size}

A key challenge for multi-object integral field spectroscopy is to
obtain sufficient signal-to-noise at low surface brightness levels for
all the targets observed simultaneously in a given pointing. In this
section we present some preliminary investigations of the properties of
potential SAMI targets. We will use the recent Sersic fits and
bulge-disc decomposition carried out by \citet{2011arXiv1107.1518S} on
the SDSS photometry.

\begin{figure}
\includegraphics[width=68mm,trim = 40 30 0 200]{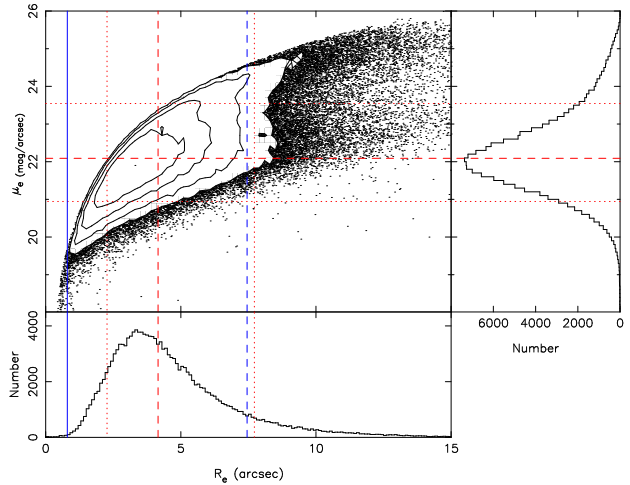}
\caption{The size--surface brightness distribution for an $r$-band
  sample of SDSS galaxies limited to $r\ser<16.5$ (extinction
  corrected). Size is defined as the half-light radius, $\re$, and the
  surface brightness, $\mue$, is also given at this radius. The marginal
  distributions of size and surface brightness are shown below and to
  the right of the main panel. Red dashed lines show the median of each
  parameter, while the dotted red lines show the 10th and 90th
  percentiles of the distributions. The blue solid line marks the radius
  of a single fibre core in SAMI and the blue dashed line is the radius
  of the 61-core hexabundle.
\label{fig:size_sb}}
\end{figure}

First we consider a simple apparent-magnitude-limited sample with
$r\ser<16.5$ (extinction corrected, where $r\ser$ is the galaxy
SDSS $r$-band magnitude derived from a Sersic model fit to the photometry). This limit
was chosen as the galaxy surface density approximately matches the
density of IFUs in SAMI. The distribution of half-light radius, $\re$,
and surface brightness at $\re$ is shown in Figure~\ref{fig:size_sb}.
For such a sample the SAMI hexabundle IFUs reach to $1\re$ for all but
the largest 10~percent of galaxies (blue dashed line), while $1\re$ is
sampled by at least 3 IFU elements for all but the smallest 10~percent
of galaxies. In other words, for the central 80~percent of this sample,
SAMI can give spatially resolved spectroscopy out to at least $1\re$.
The median surface brightness at $1\re$ is
$\mue\simeq22$\,mag\,arcsec$^2$.

\begin{figure}
\includegraphics[width=70mm,trim=20 360 20 250]{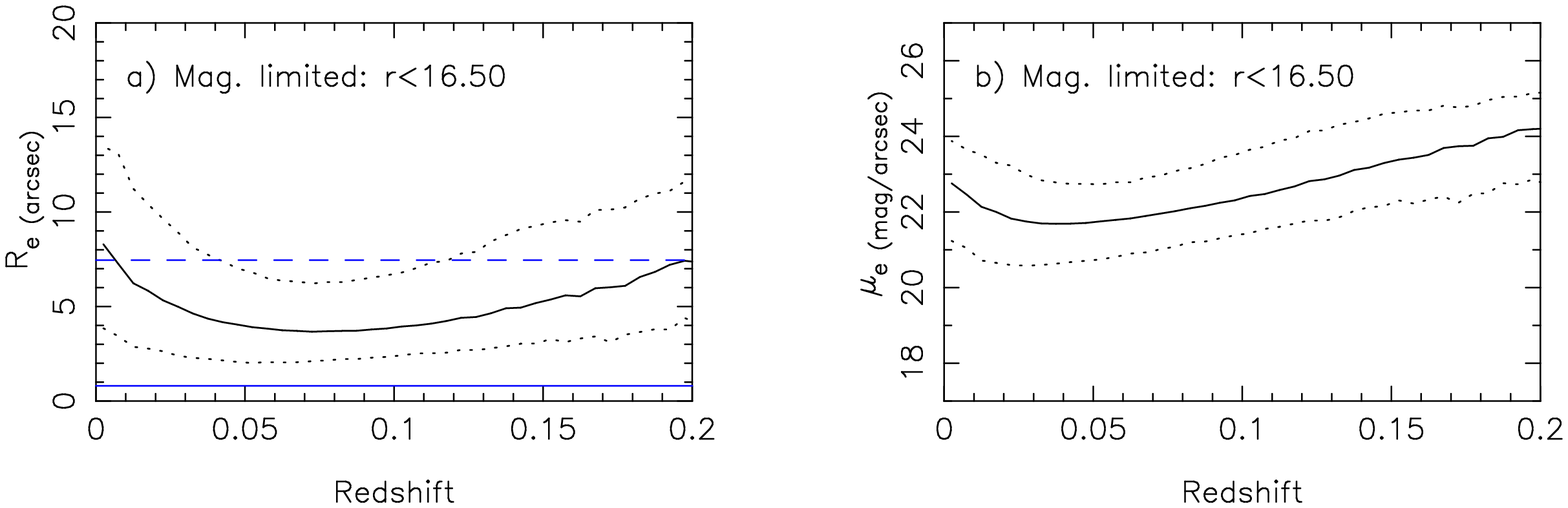}
\includegraphics[width=70mm,trim=20 320 20 200]{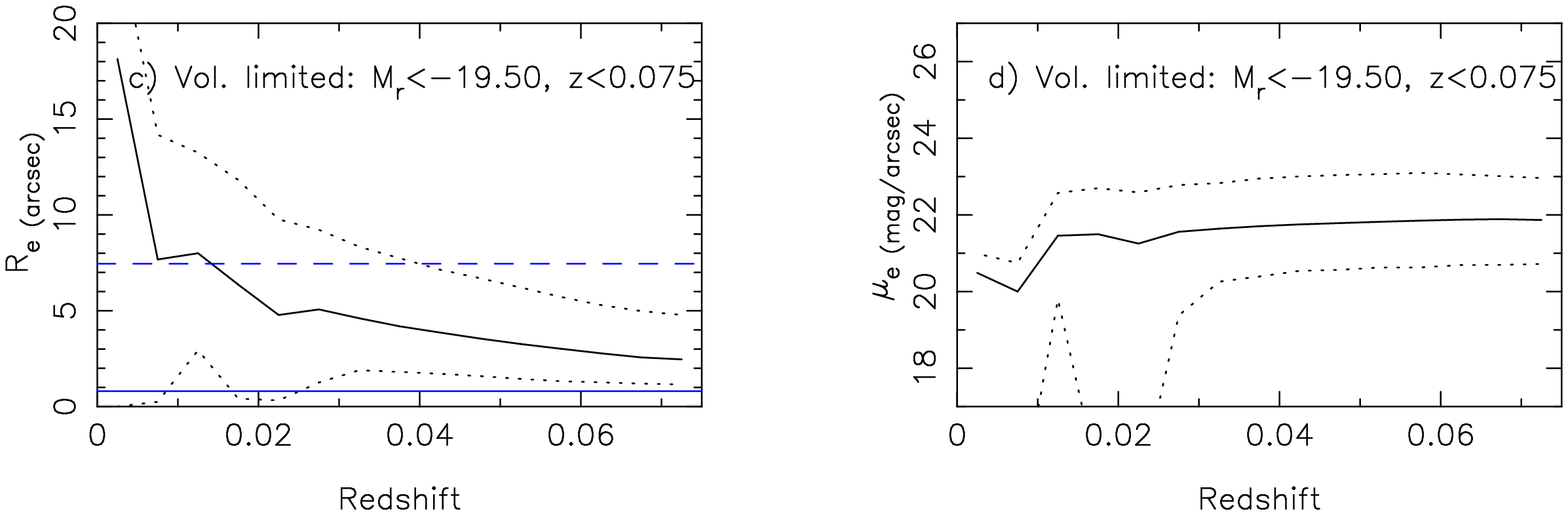}
\caption{Size and surface brightness as functions of redshift for a
  magnitude-limited galaxy sample with $r\ser<16.5$ (panels a and b) and
  for a volume-limited sample with $M_{\rm r}<-19.5$ and $z<0.075$
  (panels c and d). The black solid line is the median and the black
  dotted lines are the 10th and 90th percentiles. The
  blue solid line is the radius of a single fibre core in SAMI and the
  blue dashed line is the radius of a 61-core hexabundle.
\label{fig:size_sb_z}}
\end{figure}

We show how size and surface brightness vary with redshift for this
sample in Figures~\ref{fig:size_sb_z}a and~b. The typical sizes in
arcseconds of the galaxies stay relatively constant with redshift,
largely because the $r$-band limit selects more massive (and
therefore larger) galaxies at higher redshifts. A natural alternative is
to select a volume-limited sample, which is shown in
Figures~\ref{fig:size_sb_z}c and~d. In this case we choose 
$M_{\rm r}<-19.5$ and $z<0.075$, which gives similar numbers of targets
(i.e.\ similar surface density) to the $r$-band cut used above. In this
case we see, unsurprisingly, that the galaxies are smaller at high
redshift, but that the median $\re$ is more than twice the radius of a
fibre core (and more than three fibre cores out to $z\simeq0.06$).

An apparent-magnitude-limited sample can be considered as a set of
volume-limited samples with small redshift intervals and absolute
magnitude limits that vary with redshift. It is highly likely that the
optimal solution for targeting galaxies for a multi-object IFU
instrument involves taking multiple volume-limited samples, allowing
optimal use of the IFU FoV while broadly populating the
distribution of galaxy stellar mass. One issue that needs to be
considered is that in an apparent-magnitude-limited sample, objects
inevitably pile up around $L^*$, so suitable sampling may need to be
implemented to efficiently cover a wide range in stellar mass.

Last considerations in target selection include whether to sample all the
galaxies in a given volume or to choose field locations which uniformly
sample the distribution of galaxy environment. The challenge with the
latter approach is that there are a variety of different local density
estimators and the relation between them is not trivial (see Brough et
al., in preparation).  There is also a challenge to balance the
requirements of maintaining sufficient spatial resolution below
$1\re$, as well as reaching $>2\re$ which is a requirement for
reaching the turn-over in the rotation curves of disk galaxies
(e.g. for Tully-Fisher analysis).  Separate samples within the same
survey area that can be observed concurrently may be the most natural
solution to this problem.  Further discussion of the detailed
sample selection for a large SAMI galaxy survey will be deferred to a
future paper.

\section{The Sydney-AAO Multi-object Integral field spectrograph (SAMI)}
\label{sec:instrument} 

With a view to providing the first on-telescope demonstration of
hexabundle technology, the Australian Astronomical Observatory (AAO) and the University of Sydney have
collaborated on the development of the SAMI instrument for the
3.9m Anglo-Australian Telescope (AAT). SAMI
uses 13$\times$61-core hexabundles that are mounted on a plug-plate at
the 1-degree FoV triplet corrector top-end focus of the AAT.
At f/3.4, with 105\,$\mu$m core diameter fibres, each hexabundle samples
a 14.9\,arcsec diameter field at 1.6\,arcsec per fibre core. At the
output end, a total of 13 V-groove slit blocks are mounted at the slit
of the AAT's AAOmega spectrograph (\citet{2006SPIE.6269E..14S}; also see
Section \ref{sec:aaomega}). Each slit block includes 63 fibres (all the fibres from
one hexabundle plus two fibres for sky subtraction). A fusion-spliced
ribbonised fibre cable of length $\sim$42\,m joins the two instrument
ends together. A near real-time data pipeline, based on the {\small 2dFdr} code
\citep{2004AAONw.106...12C,2010PASA...27...91S}, has been written to reduce the data.

The following subsections describe the instrument requirements, fibre cable, the hexabundles, the
prime focus unit, the SAMI field plates, the AAOmega spectrograph, the instrument control and
data reduction software.

\subsection{Instrument requirements}

The SAMI instrument was designed to be a technology demonstrator
and to carry out significant science programmes.  As  a result, the final
instrument design is influenced by a mix of scientific and technical
constraints.  A key constraint was to develop the system on a rapid
time-scale, which naturally led to the use of the plug-plate system and
the already available AAOmega spectrograph.  The flexibility of
AAOmega, with a range of resolutions and wavelength settings also
enables a variety of science.

In order to make a substantial advance over previous facilities, the
multiplex of the system had to be at least an order of magnitude
better than previous monolithic IFUs (i.e. SAMI required at least
$\simeq 10$ IFUs).  The median optical seeing at the AAT is $\simeq1.5$
arcsec, so the fibre cores were approximately matched to this (1.6
arcsec diameter).  Although undersampling 
the seeing, this is generally preferable to having smaller core sizes for a
number of reasons: i) larger fibre cores provide more independent
resolution elements, ii) smaller fibre cores will lead to data being
read-noise limited in the blue, iii) larger fibre cores provide better
surface brightness sensitivity, iv) larger fibre cores allow higher
fill factors, given a fixed minimum fibre cladding (i.e. $5\mu$m in
our case), v) critical sampling of the seeing can still be achieved with
dithered exposures.

Given the above, the final key instrument design decision was the
number of fibres per IFU.    The choice of 61 fibres per IFU was made
largely on the basis of the known capability to manufacture such
bundles.  However, this number of fibres also matches other
requirements.  Larger numbers of fibres per bundle would have
restricted the multiplex given the fixed AAOmega slit length, and the
chosen 15 arcsec IFU diameter provides a good match to the scale
length of galaxies in local samples chosen to match the surface
density of SAMI IFUs within the 1--degree diameter field of view (see
Section \ref{sec:size}).  If designed for a single experiment, one
option would have been to have a variety of bundle sizes to match the
specific galaxy size distribution of the target sample.  However, for
ease of manufacture, and to maintain multiplex and flexibility, we
chose to have all the IFUs the same size.

\subsection{Fibre cable}
\label{sec:fibres}

SAMI is mounted at prime focus on the AAT and feeds the AAOmega
spectrograph located in the Coude room, requiring a fibre cable run of
$\sim$42\,m. Within the fibre bundle a total of 819 fibres are used: 793
for the hexabundles (13$\times$61 fibres), and 26 for the sky fibres.
The hexabundles were each supplied with a fibre pigtail length of 1\,m
encased in a reinforced furcation tube. Each of the hexabundle and sky
fibres were fusion spliced onto a fibre of length $\sim$42\,m that is
terminated in a V-groove block at the spectrograph entrance slit (see
Section~\ref{sec:aaomega} below). For this fibre run we chose a fibre
ribbon cable to minimise the required assembly effort. Each 250\,micron
thick ribbon contains 8~fibres, and 8~ribbons are used for each of the
13 units (one hexabundle plus two sky fibres). To minimize losses in the
splicing process, and because of its ready availability, we matched the
ribbonised fibre type to that used in the hexabundles (ThorLabs
AFS105/125Y).

A `splice-box' is mounted on the internal wall of the telescope
top-end barrel that incorporates 13 closed-cell polyethylene trays to
secure and protect the individual splices of each unit. The inputs to
this box are the 26 sky fibres (each individually sleeved with reinforced
PVC furcation tubing and terminated with SMA connectors) and the 13
individual hexabundle tubes. The output of this box is the fibre bundle.
The bundle is protected during its run (which goes through the top end
telescope ring, beside the primary mirror support, and through the
declination drive axis to the Coude room) by an inner covering of
braided cable sleeving and an outer double-split nylon conduit giving
light-weight yet relatively strong protection.

Though the use of ribbonised fibre worked well in terms of handling and
assembly, tests showed that ribbonising caused a loss due to focal ratio
degradation (FRD), as discussed in Section~\ref{sec:frd}. The fibre type used
provides relatively good throughput above 450\,nm, though it suffers
higher absorption losses in the UV than other fibres more commonly used
in astronomical instruments.  The current fibre cable described in
this paper will be replaced in the first half of 2012 to provide much
improved blue throughput.

\subsection{Hexabundles}
\label{sec:hexabundles}

\begin{figure}
\includegraphics[width=80mm]{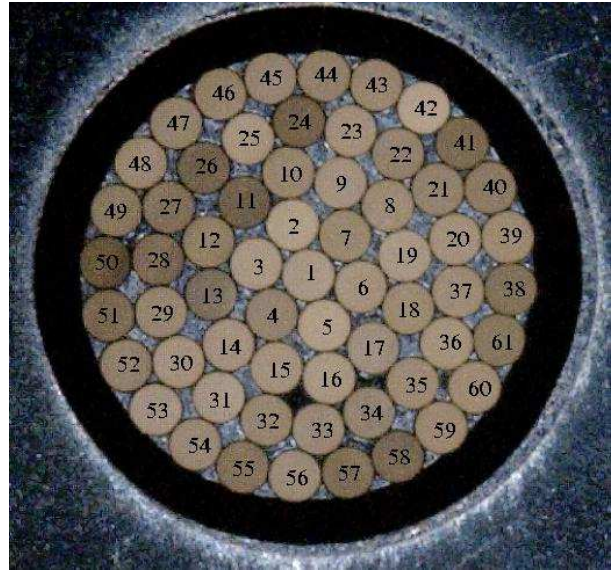}
\caption{An image of a hexabundle front facet. This shows the 61 optical
  fibre cores, which have a total active area of diameter 980\,$\mu$m.
  Surrounding the fibre cores is a glass ferrule (black ring), which
  in turn is surrounded by the central steel pin of the SMA connection,
  which extends beyond the edge of the image.
\label{hexanum}}
\end{figure}

\citet{2011OExpr..19.2649B} introduced a new imaging fibre bundle
optimized for low-light astronomical applications. The hexabundle
incorporates several technological innovations in order to achieve a
high fill factor. A primary goal of the hexabundle design is that the
performance of individual multimode fibres should be as good as the same
fibre in isolation. Consequently, we were forced to reject an earlier
design where the fibre cores were forced into non-circular shapes
because the added focal ratio degradation (numerical aperture
up-conversion) was found to lead to significant light loss
\citep{2011MNRAS.415.2173B}.

Within each hexabundle, the 61 circular fibres (with numerical aperture,
NA=0.22) are lightly fused together and infilled with low-stress glue.
The multimode fibres have relatively small core diameters of
105\,$\mu$m. In order to achieve a high filling factor (75 percent), it was
necessary to reduce the cladding to only 5\,$\mu$m. The length of the
fused region is very short ($\approx$30\,mm) to minimize cross-talk
between the fibre cores. The fibre bundle is held within a reinforced
flexible plastic tubing that is both strong and light-weight. The
optical head is supported by a stress-relieving sleeve and is inserted
into an SMA connector. This is to allow the fibre bundle to be manually
attached to the field plate with relative ease. The repeatable
positioning accuracy made possible by the SMA connector is much better
than a fibre core size.

The use of bare fibre hexabundles directly at the telescope focal
plane is the most novel element of our instrument design. There are
several advantages to such an arrangement as an alternative to
fibre-lenslet coupled IFUs that have been used elsewhere
\citep[e.g.][]{2006NewAR..50..355K}. First, the hexabundle is fabricated as a one-step process,
whereas a fibre-lenslet system requires the fibre bundle to be
manufactured and then bonded to the lenslet array(s). Secondly, there
are no optical elements required in our hexabundle design. Lenslet or
microlens array IFUs require at least 2 surfaces and more often up to
8 surfaces before the fibre face, to adjust for plate scale and to
preserve telecentricity of the telescope beam into the fibre. These
extra surfaces add to system loss. A trade-off must be made against
the potential for extra losses in the hexabundle (e.g. from FRD - see below) and from its
lower fill factor. For SAMI, as we are feeding the hexabundles with a
fast beam the FRD should be minimised (if used with the appropriate
fibre type). Finally, the hexabundle solution offers the opportunity
of reduced pitch between IFUs relative to a lenslet solution that must
work at a higher magnification. 

\subsection{Focal ratio degradation and throughput}
\label{sec:frd}

\begin{table}
\begin{center}  
\caption{FRD and throughput results for the SAMI fibres and
    hexabundles. The bare and ribbonised fibre used is the same AFS
    fibre as in the hexabundles. Results are shown for the central
    core (core 1) of hexabundle number 15 before the hexabundle was
    spliced to the fibre run. Then for core 1 and for 2 other cores (6
    and 18) in the same hexabundle, when the bundle was first spliced to
    the 42\,m of ribbonised cable with slit block attached. Lastly, for
    the hexabundle with 42\,m of ribbonised cable and slit block, but
    measured after the first SAMI commissioning run, when the ribbonised
    cables had been packed into a braided cable sleeve and outer nylon
    conduit. Cores 1 and 18 are on the edges of a ribbonised cable,
    while core 6 is in the centre of a ribbon. All measurements are
    based on an f/3.4 input to the hexabundle and f/3.15 output
    (accepted by AAOmega). The NA up-conversion is the difference between the input
    and output NA at 90 percent encircled energy and has an error in each case
    of $\pm$0.009. The
    predicted performance is what we expect to achieve when the fibre
    run has been replaced with a higher throughput fibre (e.g.\ Polymicro
    FBP) and the effects of ribbonising are removed.
\label{FRD}}
\begin{tabular}{lcccc}
\hline\hline
Fibre & \multicolumn{2}{c}{Throughput} & \multicolumn{2}{c}{NA up-conversion} \\
      & Red &  Blue & Red & Blue \\
 & \%  &  \%  & 90\% EE & 90\% EE \\
\hline
Hexabundle 15:      &          &          &       &       \\
core 1 alone        & $96\pm6$ & $89\pm5$ & 0.003 & 0.009 \\ 
core 1 plus ribbon  & $67\pm7$ & $41\pm6$ & 0.028 & 0.037 \\
core 1 plus ribbon  &          &          &       &       \\
after run           & $44\pm7$ & $24\pm6$ & 0.067 & 0.073 \\ 
 \\
core 18 plus ribbon & $64\pm8$ & $43\pm7$ & 0.026 & 0.034 \\
core 18 plus ribbon &          &          &       &       \\
after run           & $37\pm8$ & $22\pm7$ & 0.076 & 0.082 \\
 \\
core 6 plus ribbon  & $65\pm8$ & $40\pm7$ & 0.028 & 0.037 \\
core 6 plus ribbon  &          &          &       &       \\
after run           & $56\pm8$ & $32\pm7$ & 0.037 & 0.044 \\ 
\hline
Predicted performance  & 84 & 62 \\
with replacement fibre &    &    \\
\hline
42m bare fibre      & $78\pm5$ & $58\pm5$ & 0.008 & 0.018 \\
42m ribbonised      & $73\pm5$ & $53\pm5$ & 0.022 & 0.031 \\
\hline
\end{tabular}
\end{center}
\end{table}

Focal ratio degradation (FRD) increases the size of the light cone
coming out of a fibre compared to that put in. The worse the FRD, the
more light will be lost from the f/3.15 acceptance cone of the AAOmega
spectrograph. However, FRD can be partially controlled by minimising the
stresses on the fibres when installing them in the system. An additional
loss comes from the throughput of the fibres, which is primarily
dependent on the fibre type and length.  In this section we determine
the performance of the fibre cable and hexabundles by analysing FRD
and throughput.  As will be seen below, the dominant source of losses
is the fibre cable, rather than the hexabundles.
 
The FRD and throughput were measured for a SAMI hexabundle using an LED
source that was fed through Bessel $B$ and $R$ filters (centred at
approximately 435\,nm and 625\,nm) and re-imaged to form an f/3.4 beam.
This was then input into several hexabundle cores in turn. The output
fibres were imaged in the far-field using an SBIG camera. After
flat-fielding the images, the centre of each output spot was fitted.
Encircled energy was calculated in concentric circles about the centre
position using software packages within {\sc iraf}.
 
\begin{figure}
\includegraphics[width=70mm,trim=0 30 0 250]{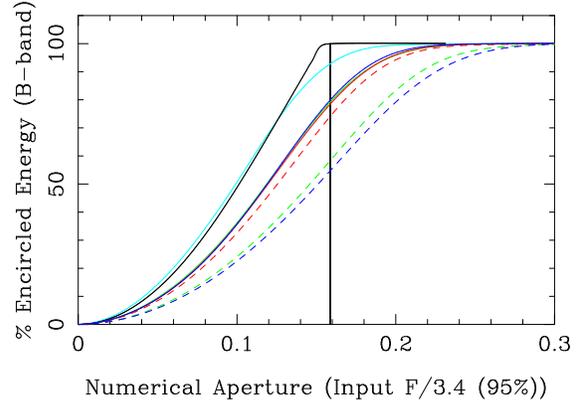}
\caption{Encircled energy (EE) vs numerical aperture (NA) profiles in
  B-band for an input of f/3.4. Larger FRD (or NA up- conversion)
  shifts the curves to the right. The vertical line marks an output of
  f/3.15 (into AAOmega). The black curve is the input light curve. The cyan line is
  the FRD of the hexabundle alone, without the ribbonised cable. The
  green, red and blue {\it solid} lines that are nearly on top of each
  other, are the curves for cores 1, 6 and 18 once spliced to the
  ribbonised 42 m cable and attached to the slit block, but before
  being put into the cable sleeve and outer nylon conduit (errors are
  $\pm0.006$ in NA). The dashed curves (with errors of $\pm 0.0065$)
  are for fibres 6, 1 and 18 (same colours as above), after the
  ribbonised fibre was put into the cable sleeve and outer nylon
  conduit and had been transported to the telescope, installed and
  used for the first commissioning run.}
\label{NAvsEE}
\end{figure}

We initially compared the performance of a hexabundle alone to the same
hexabundle when spliced to the 42\,m of ribbonised cable with a slit
block attached (hexabundle 15, `alone' vs `plus ribbon' in
Table~\ref{FRD}). For three different cores (numbers 1, 6 and 18; see
Figure~\ref{hexanum}), the throughput dropped to $\simeq65$ percent in
the red and 41 percent in the blue with the addition of the ribbonised cable plus
slit block. Figure~\ref{NAvsEE} and Table~\ref{FRD} show that this drop
is identified with a significant increase in FRD, with the NA
up-conversion (at 90 percent encircled energy) for the hexabundle with
ribbonised fibre measured to be significantly worse than for the
hexabundle alone. The
end finish on the bundle alone was a cleave, but the measurement through
the ribbonised fibre and slit block had the advantage of a polished end
finish which should improve the FRD, so the FRD introduced by the
ribbonised fibre may be a little worse than these numbers indicate.

In order to differentiate between the effect of the 42\,m of fibre and
the ribbonising, we separately compared bare and ribbonised fibre
throughputs (Table~\ref{FRD}, lower panel). This was done by measuring
the throughput from 10\,m of bare fibre of the same type as used in the
hexabundles and ribbonised fibre (AFS105/125Y). The 10\,m was then cut,
spliced together and measured to give the splice loss. Then the splice
was cut and an additional 42\,m was spliced in place. This was then cut
out and 42\,m of ribbonised cable was spliced in instead. The throughput
could then be compared for the bare and ribbonised cable after
accounting for the initial 10\,m. Having the initial fibre in place
meant that the FRD results for the bare and ribbonised fibre were not
affected by coupling into the fibre or end finishing effects as these
remained the same for both tests. Variations in throughput due to the
different splices have been taken into account in the errors.
Table~\ref{FRD} shows that while the bare fibre results will include end
effects, the ribbonising of the fibres results in significantly worse
FRD. The bulk of the loss in throughput is due to the length of fibre
(78 percent throughput in red and 58 percent in blue), however the FRD from
ribbonising results in less of that light coming out within the f/3.15
acceptance cone of the spectrograph.

Once the hexabundles were spliced to the ribbonised cable and slit
blocks (`plus ribbon' in Table~\ref{FRD}), the throughput was lower than
that measured for the 42\,m of ribbonised cable alone. This is due to an
increase in the FRD above that of the ribbonising. The small additional
FRD is therefore due to the hexabundle and slit block. Mechanical stress
on the fibres when mounted in the slit block can increase the FRD
\citep[see for example][]{2005MNRAS.356.1079O}, and hence decrease the throughput
within f/3.15.

\begin{figure}
\begin{center}
\includegraphics[width=33mm]{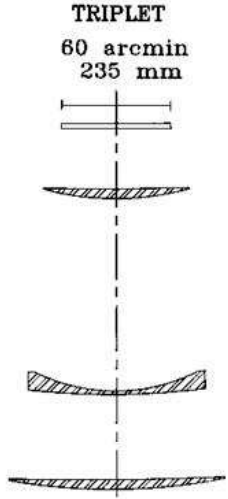}
\end{center}
\caption{Schematic of AAT prime focus triplet corrector.
\label{fig:triplet}}
\end{figure}

During the commissioning run it was noted from AAOmega images that the
throughput of the fibres on the outer edge of each ribbonised cable was
lower than that of the fibres in the centre of a ribbonised cable.
Therefore two fibres on the edge of a ribbon and one in the centre of a
ribbon were tested after SAMI had been taken off the telescope. The
testing method was identical to that described above, and the three
fibres tested were the among the same fibres from the bundle that was
tested before the SAMI commissioning run. In Figure~\ref{NAvsEE} and
Table~\ref{FRD} it is notable that while the fibres had similar
performance before commissioning (see `core plus ribbon' in the Table),
after commissioning (see `core plus ribbon after run') the FRD was
substantially worse for the two fibres on the edge of a ribbon (fibres 1
and 18). Meanwhile, the fibre in the centre of the ribbon (fibre 6)
showed only a small increase in NA up-conversion compared to the
previous results. The throughput within f/3.15 for the cores on the edge
of a ribbon decreased from 64-67 percent and 41-43 percent (red and blue
respectively) before the SAMI installation to 37-44 percent and 22-24 percent after
the SAMI commissioning run. However, the hexabundle core in the centre
of a ribbon (core 6) had a comparatively higher throughput of 56 percent and
32 percent.

We believe that the increased FRD for the edge fibres is due to stresses
in the ribbonised cable when it was fed into the braided cable sleeving
and outer nylon conduit that protects the fibre run. As the fibre bundle
was moved during transport, installation and commissioning, the
ribbonised bands could be bent in all directions, in which case the edge
fibres come under more stress than the centre.

The throughput of SAMI will soon be significantly improved by the
replacement of the fibre run. The loss due to the fibre length will be
reduced, particularly in the blue, by using a different type of fibre.
The fibres being considered will result in up to 14 percent higher throughput
at the blue end of the spectrum. In addition, if the effects of
ribbonising and FRD stresses from the ribbonised cable are removed, the
throughput would further increase by up to $\sim$27 percent, perhaps doubling
the current blue throughput. An alternative protection will be required
for the 42\,m fibre run, and that may introduce some FRD losses, but it
is being designed to have less of an effect than the ribbonising.

\subsection{Prime focus unit}
\label{sec:prime}

Within the AAT prime-focus top-end, the triplet corrector and the
Prime Focus Camera were refurbished for the SAMI instrument. Originally built for the
commissioning of the AAT in the early 1970s, they provide a 1-degree
FoV with a plate scale of 15.2\,arcsec/mm. With
61$\times$105\,$\mu$m cores, this provides a 15\,arcsec diameter FoV for
each hexabundle and a sampling of 1.6\,arcsec per fibre core (see
Figure~\ref{fig:triplet}).

With a total of 39 fibre positions in each field (26 sky positions and
13 object positions), we chose to use a plug-plate assembly rather than
a robotic positioning system, as the operational overhead and down-time
between fields for reconfiguration are both relatively low, particularly
as we are targeting long ($\sim$2-hour) integrations and have a
connectorised fibre system.

The SAMI plug plates are pre-drilled 3\,mm-thick brass discs with
through-holes at each object/sky location. Each hexabundle and sky fibre
is terminated in a SMA screw-thread fibre connector. A mating connector
is installed in each plug-plate at each position. Two galaxy fields
(i.e.\ 26 objects) are pre-drilled on the plate along with a set of 26
blank-sky locations common to both galaxy fields. For the proposed
integration time of 2\,hours per field, this means that 2~plates (and
1~plate exchange) are required each observing night. The down-time
between fields recorded during the initial commissioning run was less
than 30~minutes.  Further investigation found that it was possible to
include 4 fields per plate, so that no plate exchange is required
during the night.

\begin{figure}
\includegraphics[width=80mm]{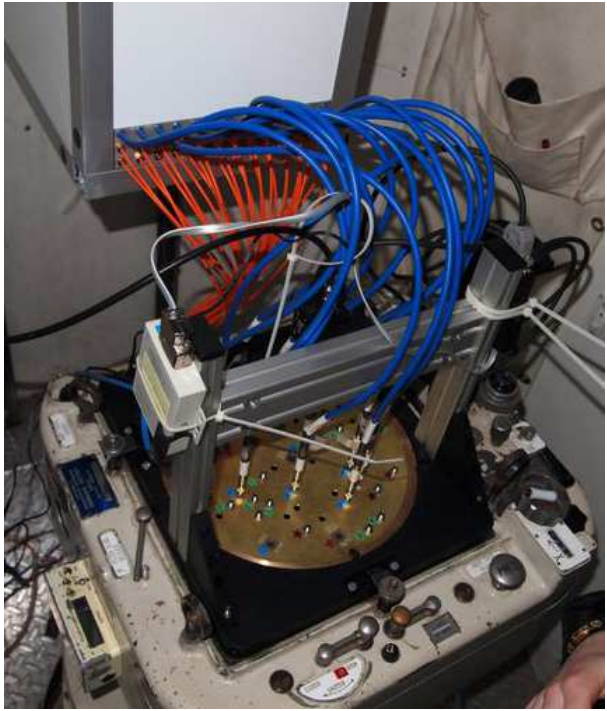}
\caption{The SAMI plug-plate assembly unit mounted onto the Prime Focus
  Camera. The white `splice box' (top) connects the blue hexabundles and
  orange sky fibres from the brass plug-plate to the fibre bundle.
  The cross-bar above the field plate provides a location to mount the
  guide camera, which images the hole in the middle of the plate.}
\label{fig:pfu}
\end{figure}

The plug-plates are installed within an assembly that is kinematically
mounted to the Prime Focus Camera (see Fig. \ref{fig:pfu}). Due to the relatively large FoV of
each hexabundle, there is not a strict requirement on the positioning
accuracy of each hexabundle central fibre. However, we aim to reach a
total positional accuracy of half the core diameter of a single fibre
(i.e.\ 0.8\,arcsec). The accuracy achieved is determined by several
factors that include: the hexabundle concentricity; the connector
concentricity; the plug-plate machining accuracy; and the plug-plate
thermal expansion. These factors are all controllable to within much
less than a fibre core. Additional position errors are introduced via
i) a rotation offset, which is corrected via a fine-thread micrometer
rotation adjustment between the plug-plate assembly and the Prime Focus
Camera, ii) an asymmetric radial error arising from an $x$--$y$ offset of
the plug plate from the telescope optical axis, for which the tolerances
are quite large and iii) a symmetric radial error arising from an incorrect
plate scale, which can be corrected after measurement by producing a new
distortion map.

For image reconstruction it is important that the rotational alignment
of each hexabundle is known. Because the SMA connector has no rotation
adjustment capability and is not rotationally keyed, the orientation of
the hexabundles was determined by eye as they were inserted into the
plug plate. Laboratory and on-sky tests demonstrated that an accuracy of
less than half a core in the outer hexabundle ring ($<$10\,degrees) was
possible with this technique. The SMA connectors will shortly be replaced by
keyed FC connectors to eliminate this limitation.

For acquisition and guiding, we use a CCD camera mounted on a gantry
above the plug-plate that views (via an optical relay) the central
region of the field through a hole in the plate. This camera (an
800$\times$600 pixel Watec \#120N) provides a video output signal that
can be integrated from frame rate up to $\sim$10~seconds and is
compatible with the existing AAT control system guiding software. It reaches
$\sim$14th magnitude over a sky FoV of 150\,arcsec
diameter (providing a sufficiently high sky coverage factor for most
galactic latitudes) and has a sampling of 0.3\,arcsec/pixel.

\subsection{Field plate manufacture}
\label{sec:plates}

Using the manual plug-plate method for positioning the hexabundles in
SAMI presents a number of challenges for the field allocation and plate
manufacturing. Here we describe the SAMI field configuration
methodology.

The plates are manufactured using brass of thickness 3\,mm and
approximate diameter 240\,mm (corresponding to the 1 degree FoV). The
fibre connectors have an assigned footprint of 15\,mm to
allow access for installation and removal. The instrument uses two
distinct physical plate types: science and calibration. Each plate is
configured with multiple stacked fields for efficiency gains. There is a
central 10\,mm hole in each plate for the guide camera to image a sky
region of diameter $\sim$150\,arcsec.

The science plates consist of two stacked galaxy fields each with 13
galaxy targets for the IFUs and about 5 field alignment stars, also
targeted with IFUs. The field alignment stars did not share the same
field centre as the galaxy targets. During commissioning, when the plate
was first observed, IFUs were positioned at the location of the
alignment stars in order to check for consistent rotation and plate
scale between each plate. Once confirmed, the IFUs were re-allocated to
galaxy targets and the telescope moved to the science field centre. Each
plate also contained 26 sky fibre positions shared between the
two galaxy fields. The science plate is configured for consecutive
fields requiring only the re-positioning of the 13 IFUs onto new galaxy
targets. Each science plate then has a total of $\sim62$ drilled holes for the
installation of the SMA fibre connectors (with a minimum separation of
15\,mm). The astrometric calibration plates were constructed in a
similar way, but in that case each plate contained four overlapping
fields and four visual alignment fields.

The process for generating a single field was as follows:\\
(i)~For each RA region, convert the RA and Dec.\ coordinates to
angular distance coordinates and calculate the pairwise distance matrix
of all targets.\\
(ii)~Iterate over each target using the distance matrix to extract
targets within a 0.5\,degree radius and to remove targets with
separations $<$15\,mm ($\sim$228\,arcsec).\\
(iii)~Count the number of targets and identify this as a candidate field
if the total is equal to or greater than $N$, where $N$ is the number of
sources required; $N$=14 for astrometric calibration fields (13 IFU
targets plus central guide star), $N$=3 for visual alignment
fields (two visual targets and a central guide star) and $N$=13 for science fields.\\
(iv)~For a candidate field, save the data and assign the field to that
RA region; eliminate the field's allocated objects from the list of
potential targets (in order to produce fields with unique sets of
targets).

Once individual fields were defined, multiple fields were stacked on a
single plate (e.g.\ for science fields the plates contain $2\times13$
galaxies plus $2\times5$ alignment fields; for astrometric calibration
fields the plates contain $4\times13$ calibration targets plus $4\times2$
visual alignment fields).  Fields were grouped into RA regions
(typically of size 20\,degrees in RA) based on observation times. The
RA regions were ordered based on the best observing time for each
region (i.e.\ the time of lowest airmass). Then fields were tested in
turn to see that the allocated targets did not overlap, so that they
could be stacked onto an individual plate. The priorities of the targets
in each science field are summed to give a figure of merit for all
possible plate permutations. The permutations can be filtered for field
duplicates and sorted by merit.

Once the plate configuration was defined, we determined the 26 sky fibre
positions that are shared for each field. This was based on a grid of
regularly spaced angular positions relative to the plate centre (e.g.\
$25\times25$ grid points over the field), which were then converted to
RA and Dec.\ given each field centre, and to relative plate
coordinates in microns (the same for each field). Sky fibre grid
locations were eliminated if they overlapped with the science targets in
plate coordinates. SuperCOSMOS images \citep{2001MNRAS.326.1279H} were examined at the RA and
Dec. of each remaining sky grid position for each field and if no
source was found within a 24\,arcsec window then the sky position was
accepted. From these remaining candidates, 26 sky fibre locations were
then selected for targeting.

For each stacked field, we then determined suitable guide stars
($V$$<$14) in the central 150\,arcsec FoV using the Aladin sky atlas
(http://aladin.u-strasbg.fr/) to search the USNO-B1 catalogue and
examine DSS images.

\begin{figure}
\includegraphics[width=95mm,trim= 0 0 0 270]{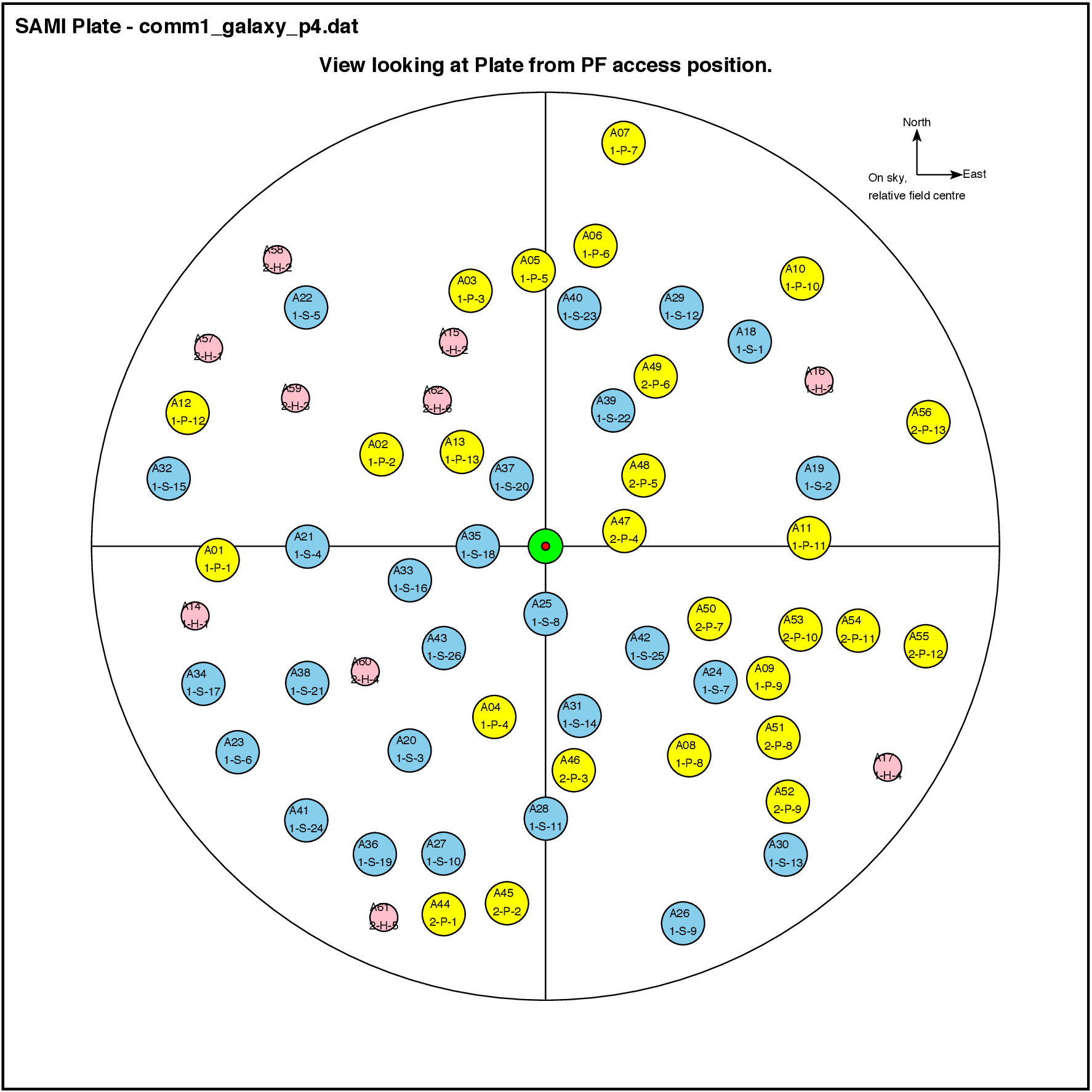}
\caption{Example SAMI science plate schematic showing 2$\times$13 galaxy
  fields (yellow/orange), 26 common sky fibres (blue), 2$\times$5 field
  alignment stars (pink) and the central guide window (green).
\label{fig:plate}}
\end{figure}

After the above process, we take the entire set of target positions in
RA and Dec.\ and convert them to accurate plate X--Y coordinates
using the optical distortion model of the prime focus corrector and the
differential atmospheric refraction at the expected time of observation.
These X--Y positions were then converted to mechanical drawings (with
temperature compensation to allow for the difference between the average
day--time drilling temperature and the average night--time observing
temperature), and these were the input to a CNC machine for drilling the
plates to the required specification.  A schematic of a galaxy field
plate is shown in Fig. \ref{fig:plate}.

\subsection{AAOmega spectrograph}
\label{sec:aaomega}

AAOmega \citep{2006SPIE.6269E..14S} is a fibre spectrograph designed for use
with the 2-degree Field (2dF) robotic fibre positioner on the AAT. It
has a double-beam Schmidt design that allows for optimised performance
in the red and blue simultaneously. Volume-phase holographic (VPH)
gratings are employed to give high throughput. There are a range of
gratings available, giving resolutions from $\sim$1700 to $\sim$13000
for the SAMI fibres (105\,$\mu$m diameter).
There is a choice of two dichroics that split the light and direct it to
the blue and red cameras; one dichroic has a cut-off wavelength of
570\,nm and a second at 670\,nm. The detectors are 2k$\times$4k E2V
devices with the short axis in the wavelength direction and the long
axis in the spatial direction.

For operation with the 2dF top-end fibre feed, the AAOmega slit is
populated with 392 fibres that each have 140\,$\mu$m cores, projecting to
$\sim$3.4\,pixels on the detector, with a core-to-core pitch of
10\,pixels. AAOmega is also fed by the SPIRAL IFU used at the AAT
Cassegrain focus. SPIRAL has a 16$\times$32 rectangular lenslet array,
with a sampling of 0.7\,arcsec. The 512 SPIRAL fibres have 85\,$\mu$m
core diameters, which project to $\sim$2.4\,pixels on the detector. The
fibre cores are separated on the slit by a pitch of 140\,$\mu$m.

For SAMI we have matched the fibre core-to-pitch ratio of SPIRAL, which
is $85/140=0.61$. This is more closely packed than used for 2dF, but has
proven sufficient with SPIRAL for the minimisation of crosstalk between
fibres using an optimal extraction methodology described by \citet{2010PASA...27...91S}. The SAMI fibre pitch on the slit is 170\,$\mu$m, giving
a core-to-pitch ratio of $105/170 = 0.62$. This pitch then defines the
maximum number of hexabundles (13) that will fit along the concave
AAOmega slit while allowing 26 individual sky fibres to also be
positioned on the slit.

For convenience of manufacture and assembly, each hexabundle and sky
fibre pair is fed
to an individual silica V-groove slitlet. The mapping is such that the
central core is in the middle of the slit.  The outward spiral
numbering of the hexabundle cores corresponds to odd and even outwardly
alternating fibres in the slitlet, with the sky fibres being the
outermost fibres on the slitlet. This arrangement maximises the number
of adjacent hexabundle cores that are also adjacent on the detector and
minimises cross-contamination between sky and adjacent object spectra.

The AAOmega slit is mounted on a rotating mechanism that allows four
different slits to be positioned in the beam of the spectrograph. This
is used during operation of 2dF, when one of the two fields is being
observed while the other is being re-configured by the robot. The SAMI
slit is located at one of the previously spare positions on the AAOmega
slit rotator; in the course of normal SAMI operations the slit does
not move.

\subsection{Control software}
\label{sec:software}

SAMI uses an adaption of the 2dF/AAOmega control software
\citep{2004SPIE.5492..410S}. This software has significant flexibility
and had been successfully evolved through various changes since it was
originally commissioned with 2dF in 1996. As part of the implementation
of AAOmega in 2006, the software was modified to support the SPIRAL IFU
feed to AAOmega. This operational mode did not use the fibre positioner
and ADC of the 2dF top-end, and as a result was a good starting point
for SAMI. A number of minor modifications were required to the software
to support this additional mode of operation.

The 2dF/AAOmega software inserts a large binary table into the CCD data
files to fully describe the location and status of each of the fibres.
This table differs significantly from versions implemented for SPIRAL
and 2dF. A component of the software (the “Fibre To Fits” program) is
used to build this table, and a new version of this program was required
for SAMI. It reads a file describing the allocation of objects to IFUs
and sky fibres. The software combines this with the telescope pointing
information and information from files containing the measured positions
of each fibre in each IFU to determine the actual location of each fibre
on the sky. As a result, a EURO-3D compliant \citep{2004AN....325..159K} binary
table can be generated and World Coordinate System (WCS) information
given for each fibre. Extra information in the table is provided to
ensure every fibre is fully traceable, and that its position and the
origins of its position are well defined.

It is important that fields be observed sufficiently close to the time
planned, and for which the plug-plate was drilled. Otherwise changes in
airmass can cause some objects to be poorly acquired, largely due to
differential atmospheric refraction across the FoV. To assist
in the decision whether to observe a given plate, the control software
displays the error between the drilled hole and the current object sky
position for each of the 13 IFU probes at the beginning of each
exposure.

\subsection{Data reduction software}
\label{sec:drsoftware}

Data reduction for SAMI is performed using the {\small 2dFdr} data reduction
pipeline \citep{2004AAONw.106...12C,2010PASA...27...91S} originally written for the 2dF instrument and
then modified for use with the AAOmega spectrograph. This provides fully
automated reduction of flat fields, arcs and object frames, including
spectral extraction, wavelength calibration and sky subtraction. The
code is run from a fully configurable GUI which allows user control of
the algorithms used. The main modifications required for SAMI were a
revised algorithm to accurately map the fibre locations across the
detector (the `tramline map') and routines to read new elements of the
FITS binary table in the data frames. {\small 2dFdr} is sufficiently fast that
data can be reduced in real time, a feature that is of particular value
during commissioning of an instrument.

The {\small 2dFdr} pipeline generates extracted, flat-fielded,
wavelength-calibrated, throughput-normalized and sky-subtracted spectra.
The data product is a 2D image containing each of the fully reduced 1D
spectra, together with a variance array and binary table (see
Section~\ref{sec:software}). To allow fast reconstruction of 3D
data-cubes and first-pass science analysis, we have developed a suite of
python-based routines which allow us to: (i)~construct and write full data cubes for individual IFUs (ii)~view all 13 reconstructed
IFU images collapsed over a user-specified wavelength range; (iii)~calculate the centroid of a source within the IFU
field; (iv)~calculate the offset in arcseconds between the source and
the centre of the IFU; (v)~extract summed spectra for an entire IFU or view individual spaxel spectra;
and (vi)~fit emission lines and construct kinematic maps on the fly.

As an aid to the development of the data reduction pipeline, an
instrument data simulator was used.  This was based on the simulator
used for the new high resolution HERMES spectrograph being developed
for the AAT \citep{2010SPIE.7735E.254G}, but modified to simulate
SAMI detector images. This was of particular value in
developing the algorithm required to extract the fibre spectra from the
2D image.

\subsection{Hardware upgrades}

\begin{table*}
\begin{center}
  \caption{Priorities used in allocating targets for the SAMI
    commissioning observations (9 is the highest priority and 1 is the
    lowest).
\label{priorities}}
\begin{tabular}{clllr}
\hline 
\hline
Priority & Input sample & Spectral criterion    & Radial size range         & Targets \\
\hline
9        & 6dFGS main   & strong emission lines & $r > 10\arcsec$          &  100 \\
8        & 6dFGS main   & strong emission lines & $5 < r \leq 10\arcsec$   &  748 \\
7        & 6dFGS-v      & high-S/N early-type   & $r > 10\arcsec$          &   28 \\
6        & 6dFGS-v      & high-S/N early-type   & $5 < r \leq 10\arcsec$   &  152 \\
5        & 6dFGS main   & strong emission lines & $4.46 < r \leq 5\arcsec$ &  107 \\
4        & 6dFGS-v      & high-S/N early-type   & $4.46 < r \leq 5\arcsec$ &  202 \\
3        & 6dFGS-v      & high-S/N early-type   & $r \leq 4.46\arcsec$     &  216 \\
2        & 6dFGS main   & strong emission lines & $r \leq 5\arcsec$        &  298 \\
1        & 6dFGS main   & all others remaining  & all others remaining      & 2235 \\
         &              &                       & {\bf Total}               & 4086 \\
\hline 
\end{tabular}
\end{center}
\end{table*}

The current system, as described above, performs well (see results
below), but substantial improvements can be made in performance and
usability.  To this end a new fibre cable will replace the current one
in the first half of 2012.  This will substantially improve the system
throughput, particularly at wavelengths $<4500$\AA.  The new fibres
will be housed in a low stress cable to further improve throughput by
substantially reducing FRD.  A third planned modification will be to
use a different form of connector for the hexabundles.  While
providing accurate alignment in the $x$ and $y$, the current SMA
connectors do not allow easy rotational alignment.  The SMA connectors
will be replaced with a keyed connector such as the FC type.

\section{Commissioning Observations}
\label{sec:comm}

\subsection{Stellar target selection}
\label{sec:starsel}

The astrometric calibration plates consist of four stacked stellar
fields ($V_{\rm T}<10$\,mag), each having 13 IFUs for distortion mapping and
alignment refinement as well as four stacked bright stellar fields
($V_{\rm T}<6.5$\,mag) each with at least two stars for initial visual alignment.
The calibration plate field uses a centred guide star located at the field
centre.

The calibration plates are derived from the stellar data contained
within the Tycho-I Reference Catalogue
\citep[TRC;][]{1998A&A...335L..65H}.  We trimmed the
catalogue on position (Dec.\,$<20$ degrees) and magnitude ($V_{\rm T}<6.5$
for visual alignment fields, $V_{\rm T}<10$ for calibration fields) for a
number of observable RA regions.

\subsection{Galaxy target selection}

The targets for the commissioning were selected to test the full
capability of SAMI, subject to the following criteria:\\
(i)~The density of targets on the sky exceeded the 13 hexabundle units
available in the 1--degree  field of SAMI.\\
(ii)~The targets were of sufficient angular size to fill the 15 arcsec
aperture of each fibre bundle.\\
(iii)~Galaxy surface brightnesses (in either emission lines or
continuum) were likely to yield S/N$>$3 in the outermost fibres of each
bundle in a $\sim$2-hour integration.

We decided to select commissioning targets from a wide cross-section of
galaxy spectral types to better constrain the performance of the
instrument across a range of future scientific applications.

The target input catalogue was drawn from the 6-degree Field Galaxy
Survey \citep[6dFGS;][]{2004MNRAS.355..747J,2005PASA...22..277J,2009MNRAS.399..683J}, which itself was
selected from the Two-Micron All-Sky Survey Extended Source Catalog \citep[2MASS XSC;][]{2000AJ....119.2498J}. The 6dFGS catalogue contains 125071
redshifted southern galaxies ($|b| > 10\degree$) selected to $K \leq
12.65$, $H \leq 12.95$, $J \leq 13.75$, $r_{\rm F} \leq 15.60$, and
$b_{\rm J} \leq 16.75$. The median redshift of the sample is $z = 0.05$.
Apparent galaxy sizes are taken from the 2MASS XSC.

The mean density of 6dFGS galaxies is $\sim$7\,$\deg^{-2}$,
substantially lower than the $\sim$19\,$\deg^{-2}$ IFU density of SAMI.
However, the low-redshift nature of the sample means that 6dFGS target
densities vary considerably over the sky, and in the densest regions,
6dFGS target densities exceed the IFU density of SAMI. By careful
pre-selection of dense regions we ensured that all 13 hexabundles were
filled for every field. Nine dense regions (each of diameter 8\,deg)
were selected over a range of hour angle and declination to provide 4086
potential targets for the final allocation of SAMI fields and IFUs.

Targets were ranked on a scale from 1 (lowest priority) to 9 (highest
priority) that was used to weight the targets for hexabundle allocation
and field placement. Targets were given a greater relative weighting if
they had: (i)~large angular size; (ii)~prominent spectral emission
lines; or (iii)~a spectrum typical of an early-type galaxy, with high
signal-to-noise.

The angular sizes used in criterion (i) were 2MASS $J$-band half-light
radii. Galaxies were divided into those with apparent radial sizes
$r>10\arcsec$ ($\log r > 1.0$), those with $5 < r \leq 10\arcsec$
($0.7 < \log r \leq 1.0$), those with $4.46 < r \leq 5\arcsec$ ($0.65 <
\log r \leq 0.7$), and the smallest ones that remained, with $r \leq
4.46\arcsec$ ($\log r \leq 0.65$). These divisions were chosen to
differentiate between those galaxies most closely matched to the
hexabundle aperture size (with galaxy diameters 10--20$\arcsec$), and
those outside this range. For the purpose of commissioning, the largest
galaxies were also given high priority to extend the range of test
subjects.

Criterion (ii) was based on matches to the emission-line galaxy spectral
templates used during the 6dFGS cross-correlation redshifting procedure
\citep[see][]{2004MNRAS.355..747J}. Note that it is not a complete sample, as
spiral galaxies which are dominated by their bulge on the scale of the
6dFGS fibre aperture (6.7$\arcsec$ diameter) may not exhibit
substantial emission lines in their 6dF spectrum.

Criterion (iii) was populated by the 6dFGS velocity
catalogue \citep[6dFGS-v;][]{2009PhDT.........1C,2011arXiv1110.1916S}, a high
signal-to-noise ($>10$ per pixel in the 6dFGS spectrum) subsample of 11288 early-type galaxies.

Selecting the target sample in this way ensured that the performance of
SAMI would be gauged across a broad range of galaxy types, while also
maximising size and surface brightness considerations, where possible.
Table~\ref{priorities} summarises the selection criteria applied to each
of the priority assignments, and the total number of available targets
in each case.

\subsection{Observations}
\label{sec:obsvns}

The first commissioning observations were carried out on the nights of
1--4 July 2011 at the AAT. The primary aims
were to test the alignment and astrometry of the plug-plate and
hexabundle units, estimate system throughput, and examine the
reliability and robustness of the instrument. The secondary goal was to
obtain the first galaxy IFU observations with SAMI to test data quality
and determine the accuracy with which physical parameters could be
extracted.

The AAOmega spectrograph setup used the 580V grating in the blue arm at
a central wavelength of 480\,nm, covering the wavelength range from
370\,nm to 570\,nm (the latter set by dichroic that splits the light
between the red and blue arms of AAOmega). With the 105\,$\mu$m fibre
cores of SAMI, the 580V grating provides a spectral resolution of
$R=1730$ (or 173\,\kms\ FWHM). The dispersion is 0.103\,nm\,pixel$^{-1}$.
This wavelength range is ideal for measuring a wide range of spectral
features in the blue parts of galaxy spectra at low redshifts ($z \sim
0.05$), such as the D4000 break, \oiii\, emission, various hydrogen Balmer lines,
and Mg\,b. In the red arm, the key spectral features are the \ha, \nii\
and \sii\ emission lines. As these are all located in a relatively
narrow wavelength range (and the target redshift range is small), it is
possible to observe all of these features using the higher resolution
1000R grating. This was set to have a central wavelength of 680\,nm,
providing a spectral range from 625\,nm to 735\,nm. The spectral
resolution in the red arm is $R=4500$ (or 67\,\kms\ FWHM), with a
dispersion of 0.057\,nm\,pixel$^{-1}$.

The first observations used astrometric calibration plates to measure
the accuracy of IFU placement, including checks of rotation, plate-scale
and distortion. An initial correction for rotation was applied by one
of us (SR) observing a small number of bright ($V_{\rm T}<6.5$) stars
by eye from within the AAT prime focus unit, using
calibration holes drilled in the plate at the position of the stars. The
stars were aligned to the holes using the rotational adjustment
micrometer on the plate holder assembly (Section~\ref{sec:prime}). Once
this first correction was made, several astrometric calibration fields
were observed in order to make precise measurements of the required
parameters.

After the astrometric observations where carried out, we targeted a
number of standard stars to measure system throughput, and then a galaxy
field was observed as part of science verification. During periods of
poor weather other tests were carried out, including examination of
guide camera flexure and variations in fibre throughput as a function of
telescope position. The detailed results of all commissioning
observations are presented in Section \ref{sec:results} below.

\section{Results}
\label{sec:results}

\subsection{Data reduction}
\label{sec:drresults}

All data presented below was reduced using {\small 2dFdr} (see Section
\ref{sec:drsoftware}).  Fibre flat field frames illuminated by a
quartz lamp are reduced first to define the location of the spectra
('tram-line map'), and to construct a fibre-flat field, which all
other frames are normalized by.  This provides correction of the
relative colour response of the fibres, but not total throughput
normalization.  Extraction of the spectra was carried out using
Gaussian fits to the fibre profiles, based on a mean profile shape.
Wavelength calibration was via an arc frames using a copper-argon
lamp.  Throughput calibration of the fibres was carried out either by
using twilight sky frames or the strength of night sky emission lines
in the object spectra.  Once the fibres were flat-fielded,
wavelength-calibrated and corrected for relative throughput, sky
subtraction was performed.  A median sky spectrum was constructed from
the 26 sky fibres.  This was then subtracted from the object spectra. 

\begin{figure}
\includegraphics[width=60mm,angle=270]{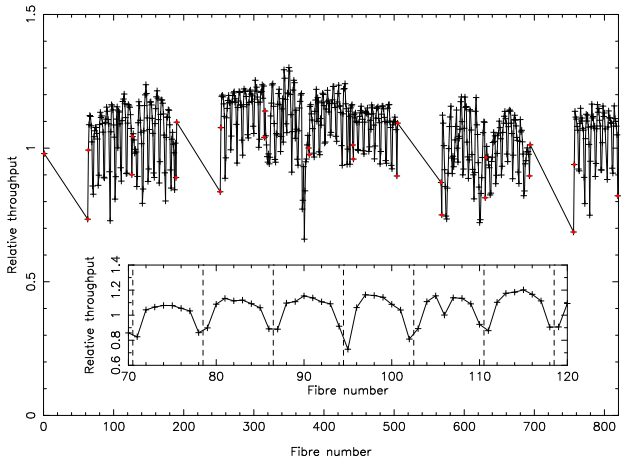}
\caption{The relative fibre-to-fibre throughput of the hexabundle
  (crosses) and sky (red circles) fibres as a function of fibre
  number.  Damaged hexabundles are not displayed (including 3 where
  the damage is so great the data is not useful, and a 4th where
  although the throughput is low, there is still reasonable S/N in the
fibres).  Inset is a small part of the throughput distribution from
fibres 70 to 120.  Here the periodic decline in throughput at the
edges of each fibre ribbon can be seen.  The dashed vertical lines
break up the fibres into separate fibre ribbons, each containing 8 fibres.}
\label{fig:fibre_thput}
\end{figure}

The relative throughput of each fibre is shown in Fig
\ref{fig:fibre_thput}.  These are normalized to the median throughput,
so are distributed about a value of 1.  The full range of throughputs
is from $\simeq0.8$ to 1.2.  A marked periodic structure is seen every
8 fibres (see inset in Fig \ref{fig:fibre_thput}) which is due to
poorer throughput at the edges of each fibre ribbon (see Section
\ref{sec:frd}) and the relative difference is consistent with the
expected throughput loss from FRD measured in the lab based
experiments.  The measured rms variation in relative fibre throughput is 0.13,
however the uniform throughput away from the ribbon edges suggests
that a rebuilt fibre cable not using ribbons would provide much more
uniform throughput.

Good throughput calibration and sky subtraction are crucial in
obtaining accurate spectroscopy at faint surface brightnesses.  Here
we make a preliminary assessment of sky subtraction precision, noting
that various upgrades to the data reduction code are still to be
implemented.  Here we focus on the residual continuum in fibres which
were located on blank sky.  Minimizing the residual continuum is
critical in enabling accurate measurements of stellar population
parameters (e.g. age and metallicity) from absorption line indices.
The  subtraction of night sky emission lines, while also important, can
be tackled in a variety of ways, including principle component
analysis \citep{2010MNRAS.408.2495S} that can effectively remove
emission to the Poisson limit of the data.  This approach will be
implemented for SAMI data.  As the galaxy commissioning targets
completely filled the field of view of each hexabundle (see
Figs. \ref{fig:gals} and \ref{fig:kin}), we used only the sky fibres
for this test.  It should be noted that examination of the spectral
point-spread function (PSF) of sky and hexabundle fibres found no
noticeable difference between them, suggesting that a test of only the
sky fibres provides a fair assessment of sky subtraction accuracy.
Future on-sky tests will confirm this with blank sky observations
using the hexabundles.  We median filter the sky spectra before and
after sky subtraction and then take the ratio of the sum of these
median filtered spectra as a measure of the fractional residual sky
continuum present.  The rms continuum residual about zero is 3.5
percent, suggesting that in the current data this is the level at
which we are accurately subtracting sky continuum.  There are a small
number (typically $\sim2$) sky fibres which perform significantly
worse that this (residuals of $\sim10$ percent or more).  Examination
of these fibres showed that i) they do not demonstrate sky emission
line residuals at the same level, and ii) they tend to be located next
to the most badly damaged hexabundle which projects essentially no
light onto the spectrograph CCD.  This points to residual scattered
light in the spectrograph as the main cause.  In particular, the fibre
profiles are made up of a Gaussian core and low level, but broad,
scattering wings.  In regions of the CCD which are fully populated
with fibres the scattering wings coadd to form an approximately
constant pedestal above the bias level of the detector (when data are
sky limited).  However, in detector regions without illuminated fibres
the background level falls below the pedestal caused by the scattering
wings.  As a result, fibres on the edge of a blank region have a
modified PSF and can be poorly extracted from the data frame.  The
solution to this is to fit a more complex PSF to the data in the
extraction process \citep[e.g.][]{2010PASA...27...91S}.  This will be
implemented within the {\small 2dFdr} package.

\subsection{Astrometry}
\label{sec:astrometry}

The AAT Prime Focus triplet corrector distortion is dominated by a
pincushion effect. For SAMI, we implemented a model of this using the
SLA library routine {\small SLA\_PCD} \citep{1994ASPC...61..481W} and
the recommended value of 178.585 for the pincushion coefficient,
$c$.  The radial distance in the presence of distortion is given
by 
\begin{equation}
\rho=r(1+cr^2)
\end{equation}
where $r$ is the radial distance from the tangent point.  Distances are in
units of the projection radius.
An analysis of a Zeemax model of the optical system agreed with the
previously used model values, to a maximum error of 0.25 arcseconds at
the field edge.  Careful attention was paid to the many orientation
and sign issues throughout the software and plate manufacturing
process, leading to the first plate being successfully acquired
with only a minor rotation correction required.

\begin{figure}
\includegraphics[width=85mm,trim=20 10 30 370]{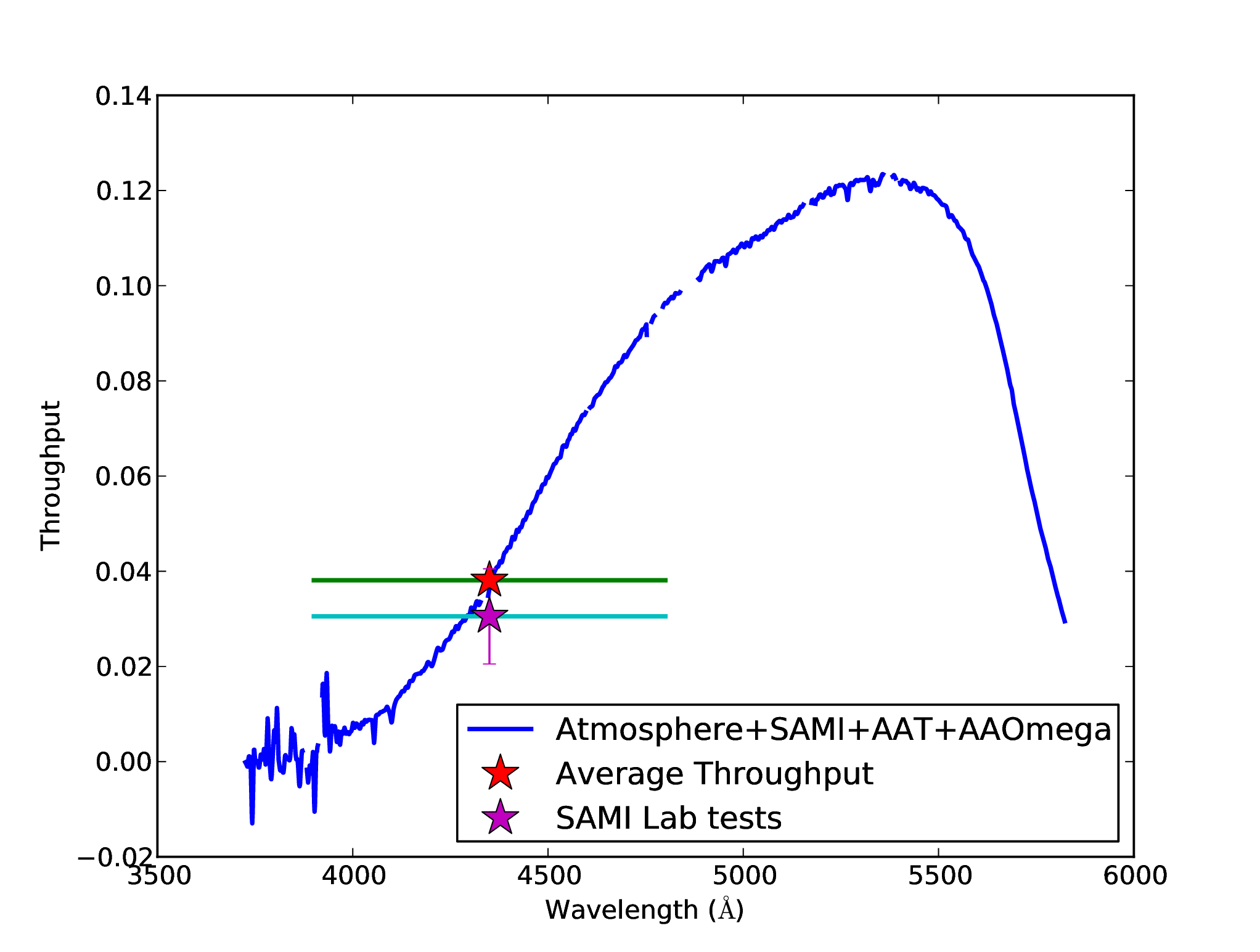}
\caption{The overall measured throughput for the SAMI system in the blue
  arm of the AAOmega spectrograph using the 580V grating. The red star
  shows the measured throughput averaged over the band indicated by the
  green horizontal line. The purple star shows the lab test estimate,
  with error bars. The horizontal cyan line
  indicates the wavelength range over which the laboratory measurements
  were made.  The low throughput in the blue was expected from our
  choice of fibre for the demonstrator instrument.}
\label{fig:tp_b}
\end{figure}

\begin{figure}
\includegraphics[width=85mm,trim=20 10 30 370]{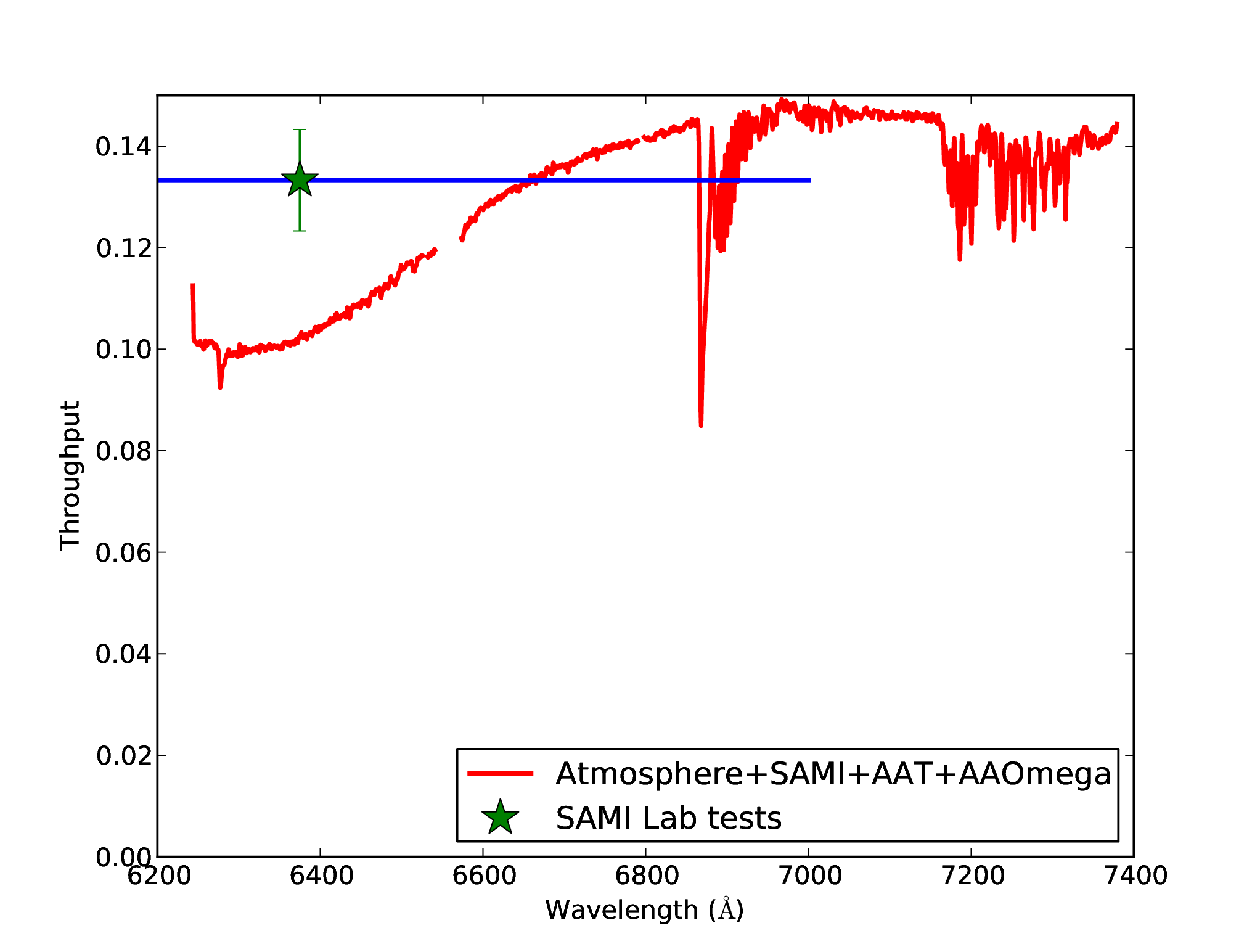}
\caption{The overall measured throughput for the SAMI system in the red
  arm of the AAOmega spectrograph using the 1000R grating. The green
  star shows lab test estimate for the throughput over the $R$-band
  (5750--7000\AA) indicated by the blue horizontal line.  The
  comparison between directly observed and lab--predicted throughputs
  is approximate as the on-sky observations did not cover the whole of
  the photometric $R$-band.}
\label{fig:tp_r}
\end{figure}

The AAO implemented a package for calibrating the astrometric models in
multi-object spectrographs for the ESO VLT FLAMES instrument 
\citep{2002Msngr.110....1P} and later
for the Subaru FMOS instrument \citep{2010PASJ...62.1135K}. This
package, known as {\small FPCAL}, 
can be configured for new instruments by the use of plug-in software
components. These components provide the program with an implementation
of the optical model and the ability to read details of objects
allocated to fibres and the telescope astrometric model parameters from
the FITS files generated by the instrument. {\small FPCAL} then provides
tools to allow fitting of model parameters and analysis of the results.
The user of {\small FPCAL} can select either Singular Value
Decomposition (SVD) or Powell's method of minimization
\cite[e.g.][]{1986nras.book.....P} as the fitting algorithm. In
practice, these give the same results within noise margins, so the
SVD technique is normally chosen as it is faster and provides more
information. The appropriate plug-ins to enable us to calibrate SAMI with
{\small FPCAL} were implemented. {\small FPCAL} requires as an
additional input the offset necessary to centre each IFU, and
software was implemented in python to extract this from the reduced data files.
Centroids were calculated using a Gaussian fit to determine the position
of calibration stars with respect to the IFU centre.

Observations of three different astrometric calibration fields were
made, including fields drilled in different physical plates. An analysis
that includes acquisition error, scale, rotation and distortion was
performed for each individual field. Sets that combine multiple
observations where the stars were offset by $\sim$2 fibre cores with
respect to the IFU centres were used to test the rotation of individual
IFUs.

In the initial analysis, results were entirely dominated by minor
acquisition and scale issues; the following refers to one typical
observation. An acquisition error of approximately 1.25$\arcsec$ was
removed to simplify further analysis. A fit to the scale found a focal
length about 20\,mm longer than originally presumed; this scale error
caused positional errors of up to 1.2$\arcsec$ at the edge of the plate.
Removing the scale error resulted in 0.8$\arcsec$ RMS residuals. Fitting
the pin-cushion distortion parameter and field rotation did not provide
a significantly better fit (0.78$\arcsec$ RMS residuals). Whilst the
change in focal length of 20mm is larger then expected, this is believed
to be due to the power of the triplet corrector, and the actual physical
change involved is likely small.

\begin{table*}
\begin{center}
\caption{Galaxies observed with SAMI during the July commissioning run.  Uncertainties on magnitudes and
  colours are typically $\sim0.1$ mags.}
\label{tab:targets}
\begin{tabular}{lcccccccccc}
\hline\hline
Hexabundle & R.A. & Dec. & Priority & 6dFGS ID & Redshift & $b_{\rm
  J}$ & 
$b_{\rm J}-r_{\rm F}$ & $b_{\rm J}-K$ & Radius$^1$ & Throughput \\
 & (J2000) & (J2000) & & & & & & & (arcsec) & flag$^2$\\ 
\hline
H\#016 & 19\,55\,18.8 & -54\,59\,35 & 1 & g1955188-545935 & 0.019 &15.43 &1.29 & 4.68 & 11.6 & good \\
H\#015 & 19\,55\,26.8 & -55\,09\,20 & 3 & g1955268-550920 & 0.045 &16.49 &1.43 & 4.30 &  3.4 & good \\
H\#005 & 19\,56\,51.4 & -54\,58\,37 & 1 & g1956514-545837 & 0.061 &16.65 &1.37 & 4.35 &  5.2 & bad \\
H\#010 & 19\,56\,51.6 & -55\,19\,56 & 1 & g1956516-551956 & 0.056 &16.35 &1.23 & 4.23 &  5.3 & good \\
H\#004 & 19\,56\,56.7 & -55\,47\,30 & 1 & g1956567-554730 & 0.018 &14.67 &1.03 & 3.62 &  8.1 & good \\
H\#014 & 19\,57\,11.2 & -55\,25\,09 & 1 & g1957112-552509 & 0.059& 16.40 & 1.33 & 4.41 &  4.4 & good \\
H\#009 & 19\,57\,22.2 & -55\,08\,14 & 1 & g1957222-550814 & 0.016 &15.36 &1.25 & 4.31 & 10.6 & good \\
H\#011 & 19\,57\,33.7 & -55\,34\,41 & 1 & g1957337-553441 & 0.017 &14.45 &1.26 & 4.32 &  8.1 & good \\
H\#001 & 19\,58\,00.3 & -55\,33\,29 & 1 & g1958003-553329 & 0.056  &16.90& 1.46 & 4.55 &  3.4 & good \\
H\#012 & 19\,58\,12.8 & -55\,40\,53 & 9 & g1958128-554053 & 0.017 &15.62 &0.16 & 4.07 & 14.5 & good \\
H\#006 & 19\,58\,29.1 & -55\,09\,13 & 1 & g1958291-550913 & 0.058 &16.14 &1.40 & 4.65 &  4.9 & bad \\
H\#013 & 19\,58\,45.0 & -55\,35\,11 & 1 & g1958450-553511 & 0.058 &16.11 &1.24 & 4.20 &  5.5 & good \\
H\#002 & 19\,56\,17.6 & -55\,07\,11 & 1 & g1956176-550711 & 0.060 &16.96 &1.73 & 3.97 &  2.4 & bad \\
\hline 
\multicolumn{10}{l}{1. $J$-band effective radii from 2MASS XSC.} \\
\multicolumn{10}{l}{2. The `bad' flag indicates the three hexabundles
  that were damaged during the run.} \\
\end{tabular}
\end{center}
\end{table*}

\subsection{Throughput}
\label{sec:throughput}

During the SAMI commissioning run two spectrophotometric standards were
observed over two nights in several of the individual IFUs. On 2~July
2011 the star LTT6248 was observed in IFU hexabundles 10 and 13 (H\#010
and H\#013). The first of these observations was used to calculate the
overall throughput of the SAMI system including the atmosphere,
telescope primary, prime focus corrector, SAMI fibre feed, and AAOmega
spectrograph. First an integrated spectrum was extracted from the IFU data by
summing the spectra in all spatial elements of the IFU. Since the data
are under-sampled and the IFU grid is not contiguous, some light is lost
between spaxels and a summed spectrum can under-estimate the true flux
in the star by as much as 10 percent. This can be avoided by performing a more
careful PSF extraction of the star over the
relevant FoV, but in this case we used the simpler method and
this should be borne in mind for the following results. The summed
spectrum was then compared to the tabulated values for the
spectrophotometric standard.

\begin{figure*}
\includegraphics[width=180mm,trim=0 20 0 0]{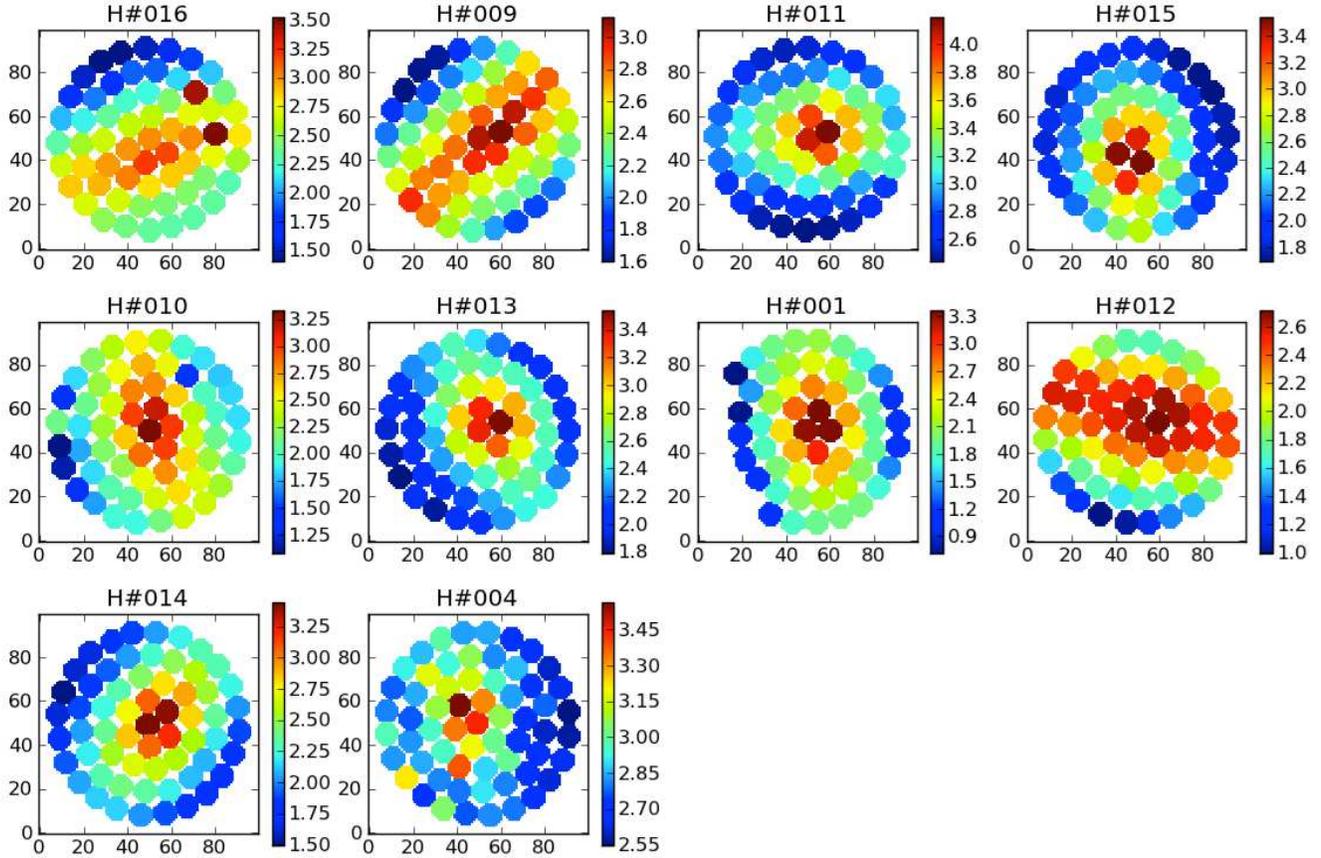}
\caption{IFU images of the observed SAMI galaxies obtained by summing
  over the spectrum in each spaxel. The X and Y positions are in
  arbitrary pixel coordinates. Each spaxel is represented by a circle
  with a 10-pixel diameter corresponding to 1.8$\arcsec$. The colour
  scales indicate the logarithm of the flux in arbitrary units.}
\label{fig:gals}
\end{figure*}

The results of this analysis are shown in Figures~\ref{fig:tp_b}
and~\ref{fig:tp_r} for the blue and red arms of the spectrograph
respectively. These throughput curves have not been corrected for
telluric features, which are still visible. The curves therefore show
the measured throughput of the entire system from atmosphere to
detector, including the AAT and the AAOmega spectrograph. Of note is the
fact that the throughput falls off quite quickly in the blue arm. This
is not ideal but the source of the effect is known. The fibre train for
SAMI consists of $\sim$42\,m of AFS105/125Y fibre matched to the fibre used in
the manufacture of the hexabundles themselves. This type of fibre has
poorer throughput in the blue than the Polymicro FBP fibre used, for
example, in 2dF \citep{2004SPIE.5492..410S}.

Throughput values measured from the standard star were compared with the
lab-tested throughput from Section~\ref{sec:frd}. This is a difficult
comparison because the lab tests use Bessel $B$-band and $R$-band filters and the
throughput measured from the standard star observations changes rapidly through the bandpass of
the Bessel $B$-band filter.  The lab tests were corrected for the known
throughput in the $B$- and $R$-bands of the atmosphere (0.72, 0.89 for $B$
and $R$ respectively),
telescope including corrector (0.77, 0.80), and spectrograph including
CCD (0.17, 0.33).  The lab tests have uncertainties due to
fibre-to-fibre differences. The data reduction includes fibre-to-fibre
throughput corrections for the standard stars, which scales the
throughput of each individual fibre to be closer to that of the fibre in the centre of a
ribbon. We therefore compare the throughput measured from the observations to core~6 in Table~\ref{FRD}.
The SAMI standard star images had a steep variation in observed
throughput across the $B$-band ranging from 26--46 percent (for SAMI only), while the
lab-tests found $32\pm7$ percent. The $R$-band lab measurement is not
directly comparable to the on-sky measurement, as 
the $R$-band filter used in the lab has a peak transmission wavelength
outside the range of the AAOmega 1000R grating.   However, we show an
indicative comparison between direct and lab--based throughput
measurements in Fig. \ref{fig:tp_r}.  The throughput comparisons, while
limited by the differences in beam shape and filters used, shows that
there are no other significant losses in the fully installed system that
are not accounted for in the above analysis of throughput.

From the measured throughput of the SAMI system we can calculate the
limiting surface brightness for typical observations, assuming photon
counting errors.  Here we assume
3 hours exposure time per field and a dark sky ($V$ and $R$ band sky
brightnesses of 21.5 and 20.8 mags\,arcsec$^{-2}$).  For a
S/N of 5 \AA$^{-1}$ the estimated limiting surface
brightness is 22.9 and 22.5 mag\,arcsec$^{-2}$ for the $V$ and $R$ bands
respectively (on the Vega scale).  For a fiducial galaxy survey sample, which is flux
limited to $r<16.5$ and redshift $z<0.13$, these surface
brightnesses limits will allow the current SAMI system to achieve
S/N$\ge5$ at 1 $r_{\rm e}$ for 80 percent of galaxies.  

\subsection{Galaxy observations}
\label{sec:gals}

\begin{figure}
\includegraphics[width=85mm,trim=0 0 0 370,clip=true]{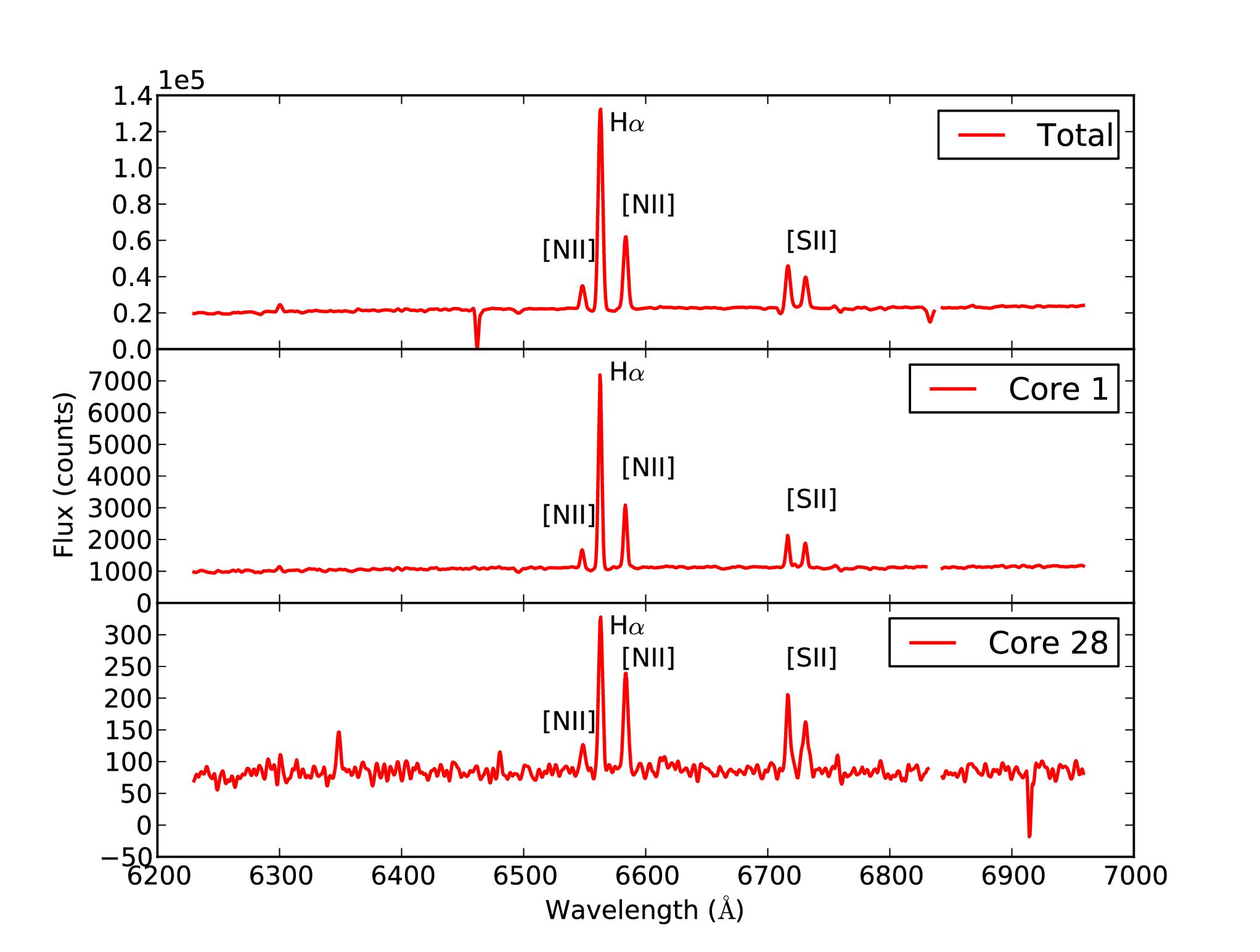}
\caption{Red spectra for the galaxy observed with bundle H\#009. The
  top plot shows the integrated spectrum converted to rest frame
  co-ordinates. The central plot shows the spectrum from core 1 of the
  hexabundle and the bottom plot shows the spectrum from core 28. The
  \ha, \nii$\lambda\lambda$6548, 6583~\AA\ and
  \sii$\lambda\lambda$6716, 6731~\AA\ emission lines are clearly
  visible.} 
\label{fig:spec_red}
\end{figure}

\begin{figure}
\includegraphics[width=85mm,trim=0 0 0 370,clip=true]{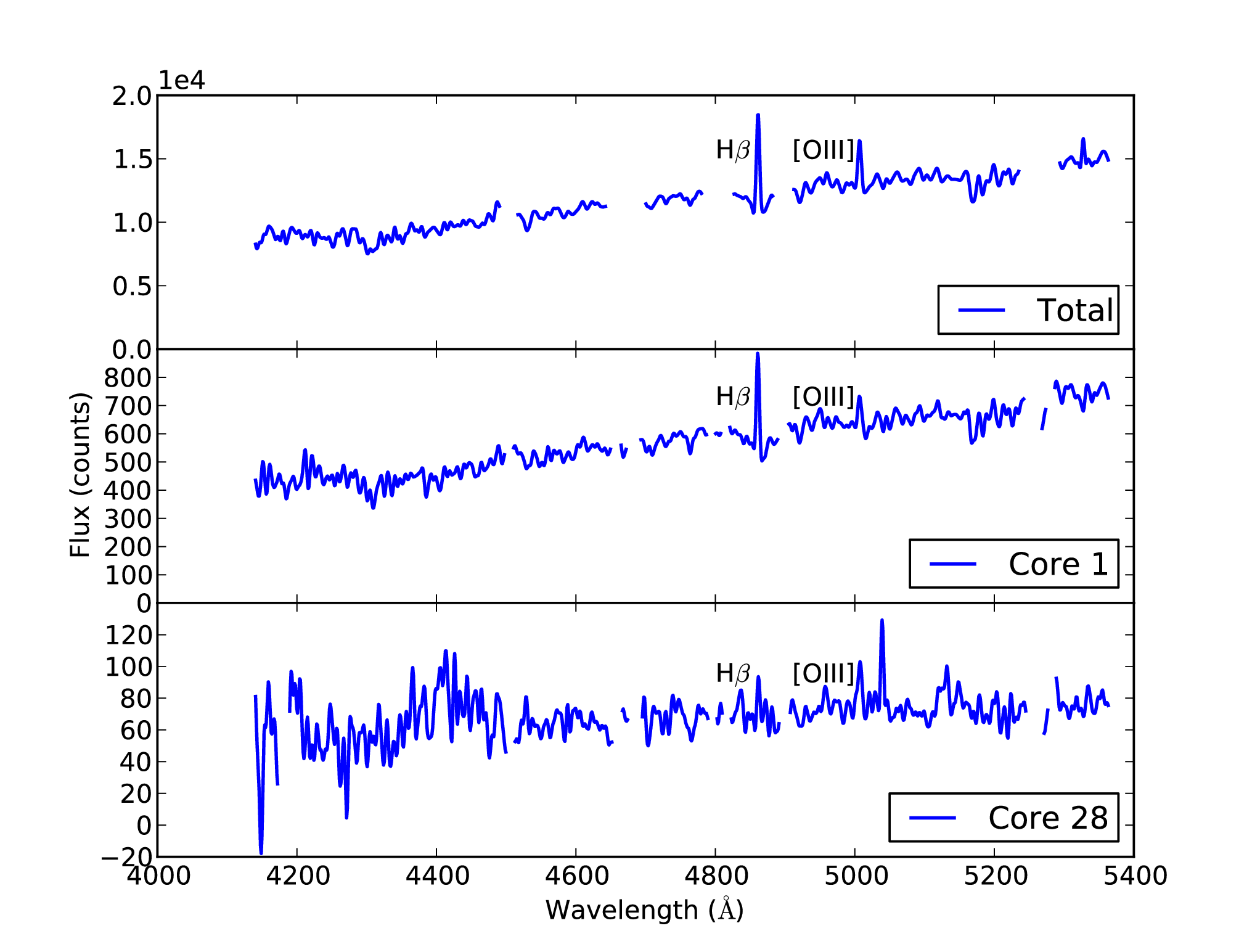}
\caption{Blue spectra for the galaxy observed with bundle H\#009. The
  top plot shows the integrated spectrum converted to rest frame
  co-ordinates. The central plot shows the spectrum from core 1 of the
  hexabundle and the bottom plot shows the spectrum from core 28. The
  \hb and \oiii$\lambda$5007~\AA\ are clearly visible in the
  integrated and central spectra but have lower S/N in the outer
  spectrum.  The masked regions in the spectra are due to bad pixels
  in the AAOmega blue CCD.} 
\label{fig:spec_blue}
\end{figure}

One galaxy field with thirteen galaxies was observed during the
commissioning run. Three 40-minute exposures were taken at this pointing
on the night of 2~July. These were reduced using {\small 2dFdr} and stacked,
resulting in a total of 2~hours integration on this field. A
spectrophotometric standard star was observed on the same night as the
objects and was reduced in an identical way. The extracted star spectrum
was then used to correct the galaxy data for instrument throughput,
though the data were not fully flux-calibrated. The details of the 13
target galaxies are given in Table~\ref{tab:targets}.

Figure~\ref{fig:gals} shows the ten galaxies from the observed field
that had usable data.   The remaining 3 hexabundles were damaged
during commissioning.  The images were created by summing over a wide
wavelength range (6300--7300~\AA\ in the observed frame); the colour
bars indicate the logarithm of the flux in arbitrary units. The discs
and bulges of the larger galaxies are readily apparent

Representative spectra of the galaxy observed in bundle H\#009 are
shown in Figures \ref{fig:spec_red} and \ref{fig:spec_blue}. The three
spectra in each figure correspond to the integrated spectrum of the
galaxy (top panel), the spectrum observed in the central core (core 1;
the central panel) and the spectrum observed in an outer core (core
28; the bottom panel). The integrated spectra are simply the sum of
all the individual spaxel spectra in a particular IFU. Various
emission and absorption features are seen in the galaxy spectra. In
the red wavelength range the \ha, \nii$\lambda\lambda$6548, 6583~\AA\
and \sii$\lambda\lambda$6716, 6731~\AA\ emission lines are clearly
visible in all three spectra, though with lower S/N in core 28. In the
blue wavelength range we see the \hb\ and \oiii$\lambda$5007~\AA\
emission lines, with an \hb\ absorption feature also visible. The data
in core 28 have lower S/N than the central core, but still sufficient
for analysis. 

\begin{figure}
\includegraphics[width=85mm,trim=0 0 0 350]{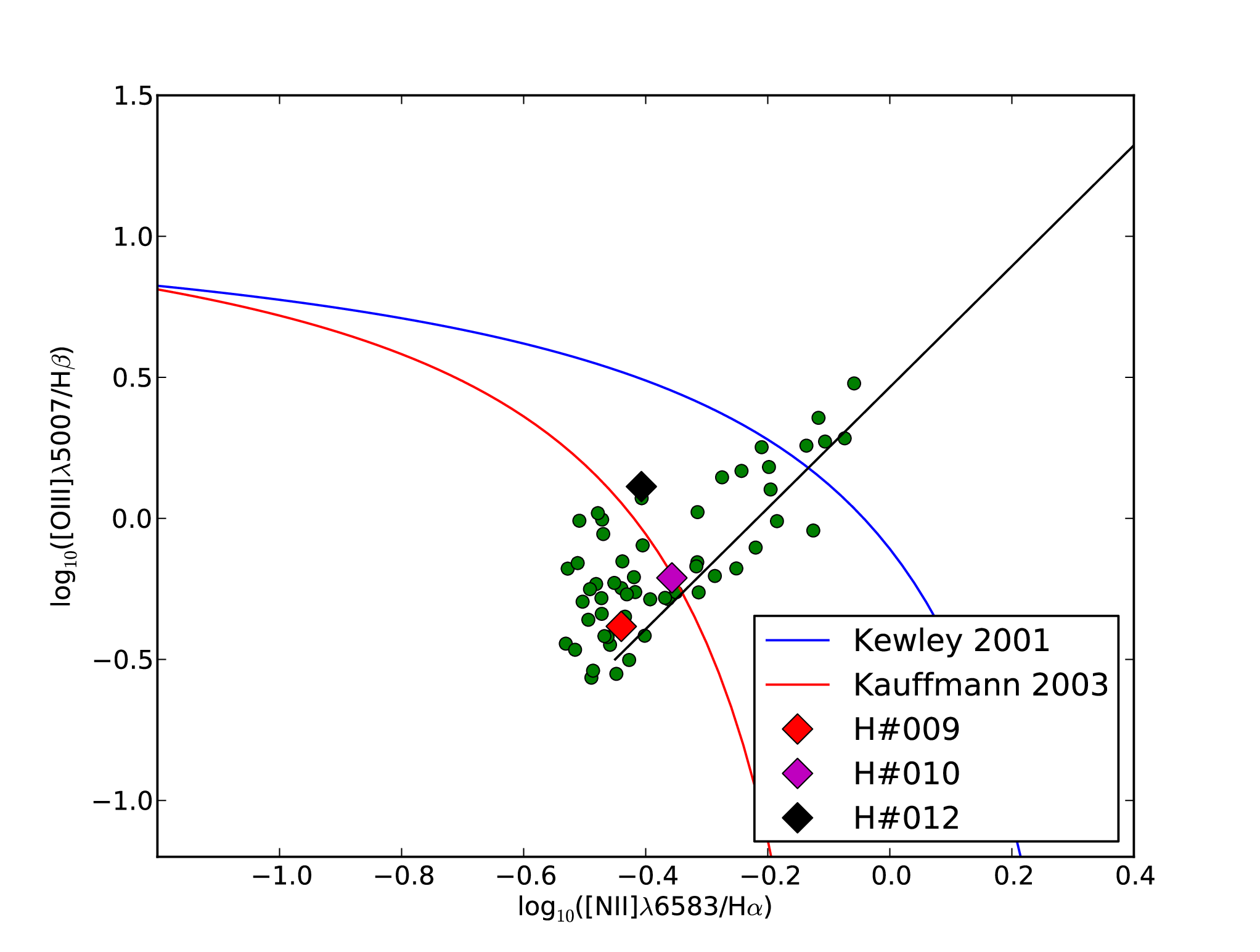}
\caption{A BPT diagram showing the positions of the three SAMI galaxies
  for which it was possible to fit all lines. Also shown are the
  theoretical maximum for star-forming galaxies (blue curve) and the
  empirical cut-off for star-forming galaxies (red curve); AGN and
  LINERs are separated, empirically, by the black line (AGN above,
  LINERs below); see text for details. The green dots show the
  positions of each of the individual spaxels for the galaxy observed
  with bundle H\#009. 
\label{fig:bpt_all}}
\end{figure}

In examining the results, we first studied the integrated spectrum of
each galaxy. Seven galaxies show medium to strong \ha\ and \nii\
emission (along with the \sii\ doublet at 6716 and 6731~\AA) and three
of these also show the \hb\ and \oiii\ lines.  

The ratios of these lines are often used to classify objects by means of
a so-called BPT diagram \citep{1981PASP...93....5B}. Each line was
fitted with a Gaussian to measure the line strength. In the case of
\hb, a weak absorption trough was observed in the summed spectra of
all three galaxies; this trough was fitted separately and accounted for
when calculating the \hb\ line strength.
Figure~\ref{fig:bpt_all} shows the galaxies on a BPT diagram (large diamonds). In
Figure~\ref{fig:bpt_all} the blue curve shows the theoretical maximum
line for star-forming galaxies after \citet{2001ApJ...556..121K} while
the red curve shows the empirical cut-off for star-forming galaxies from
\citet{2003MNRAS.346.1055K}. The black line is an empirical line, also
from \citet{2003MNRAS.346.1055K}, that separates AGN (above the line)
from LINERs (below the line).

\begin{figure*}
\includegraphics[width=160mm,trim=0 0 0 0]{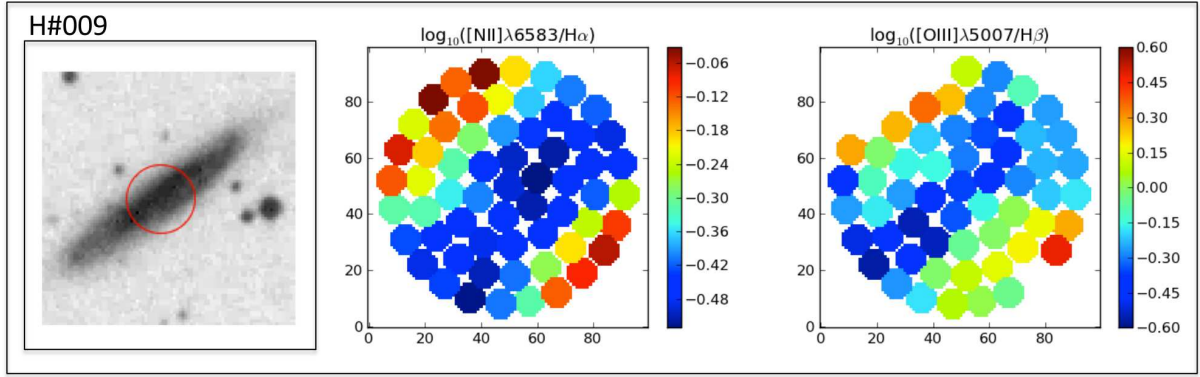}
\caption{Analysis of the spatially-resolved spectra of the galaxy
  observed with H\#009. The left-most panel shows the field of the
  hexabundle (red circle, diameter 15\,arcsec) superimposed on a UKST
  $\bj$-band image from SuperCOSMOS
  \citep{2001MNRAS.326.1279H}; the image is $1\times1$ arcminute in size. The
  other two panels show the spatially-resolved line ratio maps for the
  galaxy, with \nii$\lambda$6583/\ha\ in the centre and
  \oiii$\lambda$5007/H$\beta$ to the right.
\label{fig:h9_all}}
\end{figure*}

The galaxy H\#009 (an image of which is shown in the left-hand panel of
Figure~\ref{fig:h9_all}) was then studied in more detail, using the full
spatially-resolved set of spectra. The line ratios discussed above were
calculated for each individual spaxel in the IFU FoV, yielding
spatially-resolved line ratio maps, which are shown in the centre and right panels of Figure~\ref{fig:h9_all}. There is a systematic spatial trend
in both line ratios moving away from the plane of the galaxy's disc. The
green circles in Figure~\ref{fig:bpt_all} shows the line ratios from
each spaxel plotted on a BPT diagram. A strong trend is seen in the
sense that the spaxels further from the plane of the disc have a harder
ionising mechanism, reflecting the spatial trend seen in the line ratio
maps. A single-fibre spectrum of this galaxy would have seen only the
central few arcseconds of the object and resulted in classification as a
normal star-forming galaxy. It is clear from the SAMI results, however,
that there is a more interesting story behind this galaxy, which will be
discussed in detail in a future paper (Fogarty et al.\ in preparation).

\begin{figure*}
\includegraphics[width=160mm, trim= 0 0 0 0]{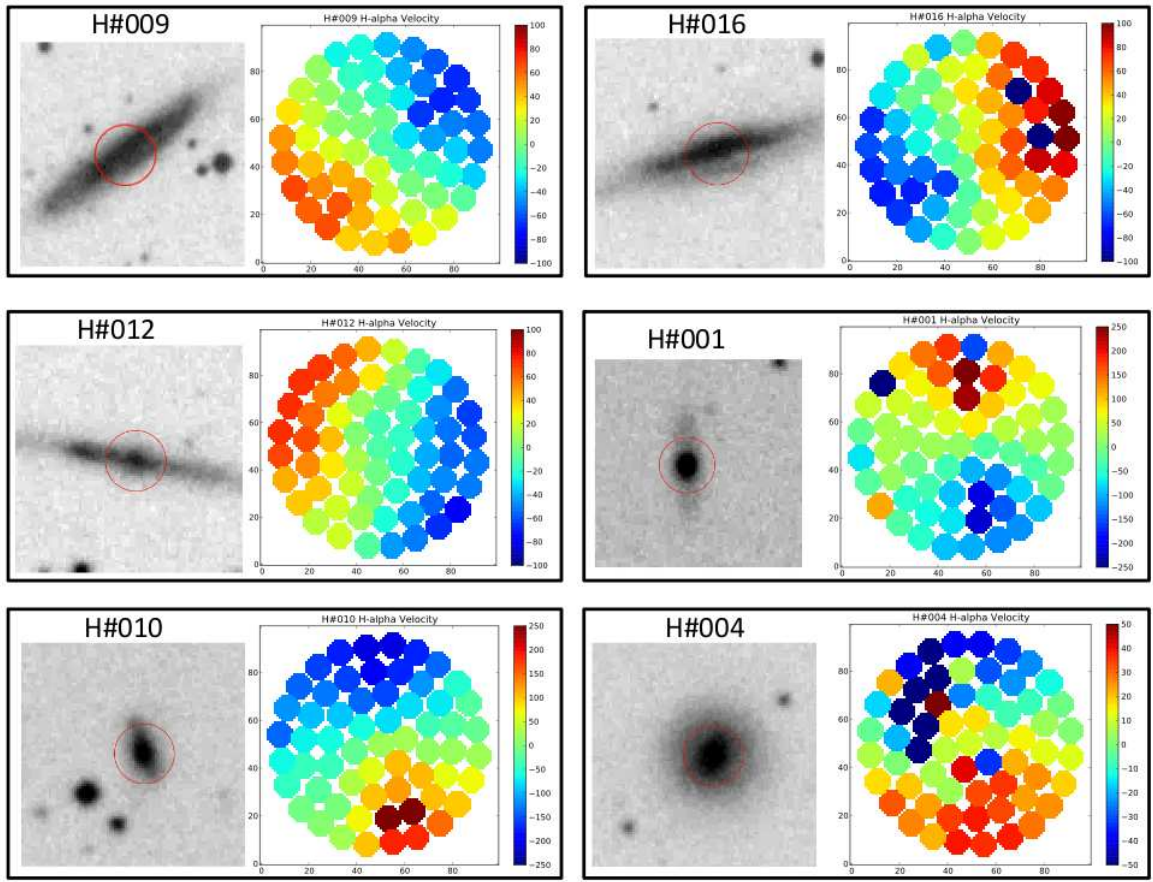}
\caption{Gas kinematics for six of the SAMI galaxies. On the left is a
  SuperCOSMOS \citep{2001MNRAS.326.1279H} $\bj$-band image of each
  galaxy, with the IFU FoV represented by the red circle. Each
  image is $\sim$1\,arcmin on a side. The SAMI H$\alpha$ velocity maps
  are shown on the right hand side, with the X and Y coordinates in
  pixel units.
\label{fig:kin}}
\end{figure*}

Gas kinematics were examined for six of the galaxies using the H$\alpha$
emission line. In each spaxel of each galaxy the H$\alpha$ line was
fitted by a Gaussian function and the kinematic properties (mean
velocity and velocity dispersion) extracted from the fit. The resulting
velocity maps are shown in Figure~\ref{fig:kin}. The colour bars show
velocity in \kms. From these maps it is clear the galaxies H\#009,
H\#010, H\#012 and H\#016 are fairly typical rotating discs; the
velocity maps for H\#001 and H\#004 are possibly more complex. 

From the wealth of spatially-resolved information presented for the
handful of galaxies studied in this single field during the
commissioning run, it is clear that SAMI can explore a very wide
parameter space with enormous science potential.

\section{Conclusions}
\label{sec:conc}

In this paper we have presented a new instrument, the Sydney-AAO
Multi-object IFS (SAMI). SAMI makes use of astrophotonic technology in
the form of hexabundles (multi-core fibre bundles) to enable
simultaneous IFU observations of 13 objects over a 1--degree diameter
accessible field. Each IFU contains 61 elements, each 1.6 arcsec in
diameter, giving a FoV for each IFU of 15 arcsec. SAMI has
now been commissioned on the Anglo-Australian Telescope (AAT) and we
demonstrate its science potential via preliminary observations of
galaxies selected from the 6dF Galaxy Survey.

We make the case that multiplexed integral field spectroscopy is the
natural next step in galaxy surveys, which to date have been dominated
by multiplexed single-aperture observations. The extra information
gained by IFU spectroscopy, combined with the statistical power of a
large survey, will enable a fundamental step forward in our
understanding of galaxy formation and evolution by distinguishing the
spectroscopic properties of the major structural components of galaxies.
Assuming that a survey could observe 3 fields per night (i.e.\ nominal
exposure times between 2 and 3\,hours), SAMI will allow 10,000 galaxies
to be targeted in 260 clear nights on the AAT.

Much greater gains could be made with an increased FoV and
larger numbers of IFUs. A key requirement for making full use of the
larger fields of view available on current and future telescopes is to
have sufficient spectrographs to handle all the fibres from the IFUs. A
2--degree diameter FoV (e.g.\ that available with the AAT's 2dF
corrector) could accommodate $\sim$50 or more IFU bundles comparable to
those in SAMI; this would, however, require $\sim$3300 fibres (including
sky fibres). Such large numbers of fibres naturally drive designs
towards mass-produced fixed-format spectrographs, such as those
currently being developed for projects such as MUSE
\citep{2004SPIE.5492.1145B} and VIRUS \citep{2006NewAR..50..378H}.

More development work on the bundle technology is under way. One
possible improvement is to force the fibre cores into a perfect
hexagonal configuration to simplify the data analysis, particularly when
using dithering to smooth out the discrete sampling. This will not be
easy to do with our current core size while at the same time ensuring an
AR-coated polished front facet that is not overly stressed by the
hexagonal grid. With the successful commissioning on the AAT of
fibre-based OH suppression technology \citep{2004OExpr..12.5902B,2010SPIE.7735E..40E}, we can envisage whole bundles that are fully sky-suppressed
at near-infrared wavelengths. Once the near-infrared sky is rendered as
dark as the optical sky, exposure times in both regimes are expected to
be the same and it makes sense to run optical and infrared
spectrographs simultaneously.

\section*{ACKNOWLEDGEMENTS} 

We thank the staff of the Australian Astronomical Observatory and
those at the University of Sydney for their excellent support in
developing, construction and commissioning of the  SAMI instrument.

SMC acknowledges the hospitality of the Leibniz Institute for
Astrophysics in Potsdam (AIP) during the completion of this paper.  We
thank Jakob Walcher, Martin Roth, Roger Haynes and Davor Krajnovi{\'c}
for helpful discussions. 
The Centre for All-sky Astrophysics is an Australian Research Council
Centre of Excellence, funded by grant CE11E0090.  SMC acknowledges the
support of an Australian Research Council (ARC) 
QEII Fellowship (DP0666615), an Australian Research Council Future
Fellowship (FT100100457) and an J G Russell Award from the Australian
Academy of Science. JBH is supported by a Federation Fellowship from
the ARC. CQT gratefully acknowledges support by the National Science
Foundation Graduate Research Fellowship under Grant No. DGE-1035963.


\end{document}